\documentclass[manuscript]{acmart}

\PassOptionsToPackage{dvipsnames,svgnames}{xcolor}
\usepackage[dvipsnames,svgnames]{xcolor}
\definecolor{BurntOrange}{RGB}{204, 85, 0}
\usepackage{array}
\usepackage{amsmath}
\usepackage{ragged2e}
\usepackage{enumerate}
\usepackage{natbib}
\usepackage{tabularray}
\usepackage{enumitem}
\usepackage{subfig}
\usepackage{graphicx}
\usepackage{caption}
\usepackage{makecell}
\usepackage{soul}
\usepackage{listings}
\usepackage{multirow}
\usepackage[normalem]{ulem}
\usepackage{tcolorbox}
\usepackage{booktabs}
\usepackage{pifont}
\usepackage{xspace}
\usepackage{epigraph} 
\usepackage[flushleft]{threeparttable}
\usepackage{url}
\definecolor{backcolour}{rgb}{0.95,0.95,0.92}
\usepackage{algorithm2e}
\UseRawInputEncoding
\usepackage[colorinlistoftodos]{todonotes}
\usepackage{listings}
\usepackage{comment}
\usepackage{pifont}
\usepackage{bbding}
\usepackage[figuresleft]{rotating}
\usepackage{framed}
\usepackage[strict]{changepage}
\usepackage{adjustbox}
\definecolor{awesome}{rgb}{0.0, 0.2, 0.6}
\usepackage{pgfplots}
\usepackage{makecell}
\usepackage{graphicx,multirow}
\usepackage{fontawesome}

\newlength\WIDTHOFBAR
\def\blackwhitebar#1{
  #1 {\color{blue}\rule{#1cm}{8pt}}{\color{black!30}\rule{\WIDTHOFBAR - #1 cm}{8pt}}}
\soulregister\cite7
\soulregister\citep7
\soulregister\citet7
\soulregister\ref7
\soulregister\pageref7

\definecolor{gray(x11gray)}{rgb}{0.75, 0.75, 0.75}
 	
\def\mybar#1{
  {\color{blue}\rule{#1cm}{8pt}}}

\newlength\MAX  

\newlength\Base  

\definecolor{formalshade}{rgb}{0.95,0.96,0.96}
\definecolor{side}{rgb}{0.0,0.2,0.6}

\DeclareMathOperator*{\argmax}{arg\,max}

\newcommand{\AdaBracket}[1]{\left(#1\right)}
\newcommand{\AdaRectBracket}[1]{\left[#1\right]}

\newcommand{\expectation}[2]{\mathbb{E}_{#1}\AdaRectBracket{#2}}

\acmJournal{JACM}
\acmArticle{111}
\AtBeginDocument{
  \providecommand\BibTeX{{
    \normalfont B\kern-0.5em{\scshape i\kern-0.25em b}\kern-0.8em\TeX}}}

\setcopyright{acmcopyright}

\begin{document}

\title{A Survey of Reinforcement Learning for Software Engineering}

\author{Dong Wang}
\affiliation{%
  \institution{College of Intelligence and Computing, Tianjin University}
  \city{Tianjin}
  \country{China}
}
\email{dong_w@tju.edu.cn}

\author{Hanmo You}
\authornote{Indicates Joint First Authors.}
\affiliation{%
  \institution{College of Intelligence and Computing, Tianjin University}
  \city{Tianjin}
  \country{China}
}
\email{youhanmo@tju.edu.cn}

\author{Lingwei Zhu}
\affiliation{%
  \institution{The University of Tokyo}
  \city{Tokyo}
  \country{Japan}
}
\email{lingwei4@ualberta.ca}

\author{Kaiwei Lin}
\affiliation{%
  \institution{College of Intelligence and Computing, Tianjin University}
  \city{Tianjin}
  \country{China}
}
\email{kwhy200318@gmail.com}

\author{Zheng Chen}
\affiliation{%
  \institution{SANKEN, The University of Osaka}
  \city{Osaka}
  \country{Japan}
}
\email{chenz@sanken.osaka-u.ac.jp}

\author{Chen Yang}
\affiliation{%
  \institution{College of Intelligence and Computing, Tianjin University}
  \city{Tianjin}
  \country{China}
}
\email{yangchenyc@tju.edu.cn}

\author{Junji Yu}
\affiliation{%
  \institution{College of Intelligence and Computing, Tianjin University}
  \city{Tianjin}
  \country{China}
}
\email{junjiyu@tju.edu.cn}

\author{Zan Wang}
\affiliation{%
  \institution{College of Intelligence and Computing, Tianjin University}
  \city{Tianjin}
  \country{China}
}
\email{wangzan@tju.edu.cn}

\author{Junjie Chen}
\authornote{Corresponding Author}
\affiliation{%
  \institution{College of Intelligence and Computing, Tianjin University}
  \city{Tianjin}
  \country{China}
}
\email{junjiechen@tju.edu.cn}

\renewcommand{\shortauthors}{}

\begin{abstract}
Reinforcement Learning (RL) has emerged as a powerful paradigm for sequential decision-making and has attracted growing interest across various domains, particularly following the advent of Deep Reinforcement Learning (DRL) in 2015. 
Simultaneously, the rapid advancement of Large Language Models (LLMs) has further fueled interest in integrating RL with LLMs to enable more adaptive and intelligent systems. 
In the field of software engineering (SE), the increasing complexity of systems and the rising demand for automation have motivated researchers to apply RL to a broad range of tasks, from software design and development to quality assurance and maintenance. 
Despite growing research in RL-for-SE, there remains a lack of a comprehensive and systematic survey of this evolving field. To address this gap, we reviewed 115 peer-reviewed studies published across 22 premier SE venues since the introduction of DRL.
We conducted a comprehensive analysis of publication trends, categorized SE topics and RL algorithms, and examined key factors such as dataset usage, model design and optimization, and evaluation practices. Furthermore, we identified open challenges and proposed future research directions to guide and inspire ongoing work in this evolving area.
To summarize, this survey offers the first systematic mapping of RL applications in software engineering, aiming to support both researchers and practitioners in navigating the current landscape and advancing the field.
Our artifacts are publicly available: \textcolor{blue}{\url{https://github.com/KaiWei-Lin-lanina/RL4SE}}~\faGithub.

\end{abstract}





\maketitle

\section{Introduction}
\label{introduction}
\textbf{Reinforcement Learning (RL)} is a pivotal paradigm in machine learning, where an agent learns to make decisions by interacting with an environment state, aiming to maximize cumulative rewards over time. 
The concept of RL has its roots in behavioral psychology and was formally established in the field of computer science through foundational works in the 1980s and 1990s, including Sutton and Barto's early contributions~\cite{sutton1998reinforcement, sutton1999policy, sutton1988learning}. 
Over time, RL evolved into a rigorous framework for sequential decision-making under uncertainty, where the learning problem can usually be modeled as \textit{Markov decision problems} (MDPs)~\cite{puterman2014markov}.
A major breakthrough occurred in 2015, when DeepMind successfully combined reinforcement learning with deep neural networks, training an agent capable of playing Atari games at a superhuman level, an achievement that gave rise to \textit{deep reinforcement learning} (DRL)~\cite{van2016deep}. 
This milestone, popularized by the \textit{Deep Q-Network} (DQN), demonstrated the power of integrating deep learning's feature extraction capabilities with RL's decision-making strengths.
Since then, RL has gained widespread attention and been applied to complex domains such as robotics, autonomous driving, computer vision, recommendation systems, and natural language processing (NLP).
In recent years, the rapid advancement and broad adoption of large language models (LLMs), such as ChatGPT and LLaMA, have significantly transformed the landscape of artificial intelligence. 
Researchers have increasingly explored combining LLMs with RL to enhance their adaptability and decision-making capabilities, particularly exemplified by models like DeepSeek-R1~\cite{guo2025deepseek}, which has demonstrated outstanding performance.

Spurred by the successful applications of RL algorithms in various domains, alongside the increasing complexity of modern software systems and the growing demand for intelligent automation, researchers and practitioners have been motivated to explore the use of RL in tackling a wide range of challenges in \textbf{Software Engineering (SE)}. 
RL offers a promising paradigm for building adaptive, data-driven solutions across the SE lifecycle, including software design and modeling, quality assurance, development, maintenance, and project management.
For instance, in the domain of software design and modeling, RL has been applied to tasks such as resource allocation~\cite{lakhan2022efficient}, hybrid optimization~\cite{krupitzer2022proactive}, and automated software modeling~\cite{guo2025effectively}. 
During software development, RL algorithms have been leveraged for code generation~\cite{gao2024preference}, code completion~\cite{wang2024rlcoder}, code summarization~\cite{wang2020reinforcement}, and comment generation~\cite{huang2020towards}. 
To ensure software quality, various RL approaches have been employed in software testing (e.g., test case generation~\cite{koroglu2021functional, borgarelli2024reward, giamattei2025reinforcement} and optimization~\cite{spieker2017reinforcement, qian2025reinforcement}), bug localization~\cite{chen2020enhanced}, and model verification~\cite{liu2022learning, wang2022learning}. 
In the context of software maintenance, researchers have explored the use of RL to automate code review tasks~\cite{joshi2024comparative}, support software refactoring~\cite{ahmadi2022dqn}, facilitate automated program repair~\cite{guo2024smart, hanna2025reinforcement}, and reproduce bugs~\cite{zhang2023automatically}.
Notably, recent studies have begun to explore the integration of RL and LLMs to address certain SE tasks.
For example, \citet{li2024ircoco} proposed a deep RL-based fine-tuning framework tailored for code completion. 
The framework delivers immediate rewards to detect dynamic context changes caused by continuous edits, enabling the fine-tuned language model to better capture and adapt to the evolving code context.
\citet{eom2024fuzzing} proposed a novel fuzzing technique, CovRL, which integrates coverage feedback into LLMs by leveraging the Term Frequency-Inverse Document Frequency (TF-IDF) method to construct a weighted coverage map. 
This map is then used to guide the LLM-based mutator through RL.

Although a growing number of SE studies have adopted RL, referred to as \textbf{RL-for-SE} for clarity, there remains a notable lack of a comprehensive and up-to-date survey that systematically examines how RL has been applied across diverse SE tasks.
Our comparative analysis (as presented in Section~\ref{comparative}) reveals that existing surveys primarily focus on general machine learning or deep learning alogorithms, or are limited to specific software engineering subdomains (i.e., software testing), while overlooking the distinctive aspects of reinforcement learning (RL) and its application across the broader SE landscape.
Such a survey is essential for understanding current trends, identifying research challenges, and guiding future developments in this rapidly evolving area.
To bridge this gap in the literature, we collected, reviewed, and analyzed \textbf{115} RL-for-SE studies published across 22 premier software engineering venues (including 12 conferences and 10 journals) since the introduction of deep reinforcement learning in 2015.
We then analyzed the development trends of RL in SE, summarized the key research topics, classified the various RL algorithms applied across diverse SE tasks, examined the RL construction processes described in the relevant studies, and identified the prevailing challenges and future opportunities in this field.

This paper makes the following contributions: 
\begin{itemize}
\item  (I) We are the first to present a comprehensive literature review spanning a wide SE landscape covering 115 studies that focus on the use of RL-based solutions to address SE challenges. We conducted a detailed analysis of the selected papers based on publication trends, distribution of publication venues, etc. 
\item (II) We systematically classified the research topics of RL-for-SE studies and provided a description of each primary study according to six different SE activities. 
We also conducted an analysis on these studies based on their task types, including regression, classification, ranking, and
generation tasks.
\item  (III) We classified the RL algorithms utilized for the reported SE tasks in a fine-grained manner and provided a summary
of the usage and trends of different RL categories within the SE domain.
\item (IV) We performed a comprehensive analysis on the key factors that impact the performance of DL models in SE, including dataset, model design and optimization, and model evaluation.
\item (V) We empirically summarized key challenges that RL encounters within the SE field and suggested several potential research directions for RL-for-SE.
\end{itemize}

\smallskip
\noindent
\textbf{Paper Organizations.} Section~\ref{background} provides the necessary background on RL, introducing all essential concepts relevant to this study.
Section~\ref{comparative} presents a comparative analysis of existing related work to highlight the novelty and distinct contributions of our survey.
Section~\ref{methodology} outlines the methodology of our survey, including the proposed research questions and the process used for collecting and selecting the surveyed papers.
Sections~\ref{results:trend} to~\ref{sec:chall} are dedicated to addressing our proposed research questions through detailed analysis and discussion.
Section~\ref{sec:threats} presents the main threats to the validity of this study.
Finally, we conclude our survey in Section 11.

\section{Reinforcement Learning}
\label{background}

\begin{figure}
    \centering
    \includegraphics[width=0.5\linewidth]{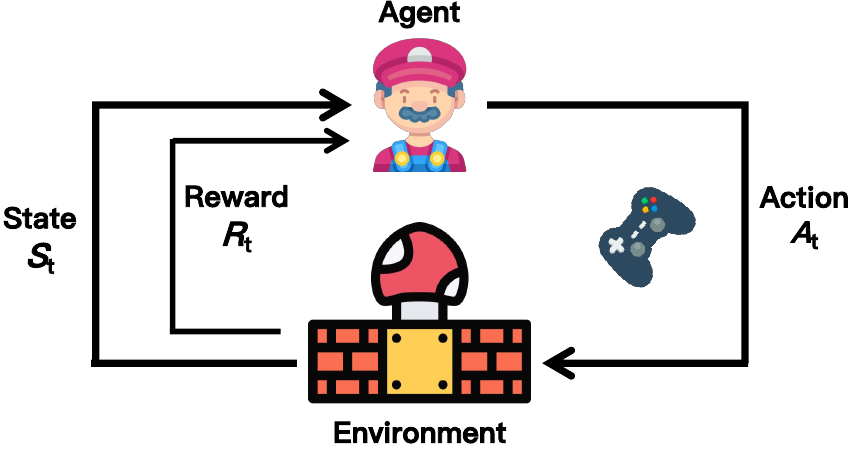}
    \caption{General framework of reinforcement learning (RL)}
    \label{fig:rl_diagram}
\end{figure}

Reinforcement learning (RL) provides a principled mathematical framework to describe how an agent can learn to perform a task by reinforcing its rewarding behaviors \cite{Sutton-RL2018}.
Different from standard machine learning branches that typically operate with independently and identically distributed samples containing explicit ground truth labels, RL centers on \emph{actions} and \emph{rewards}.
Key ingredients to an RL algorithm are states, actions, rewards, and transition dynamics.
As shown in Figure \ref{fig:rl_diagram}, at every timestep $t$, an agent recognizes the state $s_t$ it is in, and evaluates the desirability of that state, known as reward $r_t$.
Based on this information, the agent draws an action $a_t$ to cause the environment to evolve into a next state and the loop continues.
The goal of the agent is to learn an optimal policy distribution $\pi$ that dictates how actions should be chosen given states.

Formally, RL problems are formulated based on the Markov decision process (MDP) that provides a theoretical underpinning of sequential decision making  \citep{Puterman1994}.
An MDP is defined by a quintuple $\mathcal{M}=\left(\mathcal{S},\mathcal{A},\mathcal{P},\mathcal{R},\gamma\right)$, where $\mathcal{S}$ denotes the state space and $\mathcal{A}$ refers to the potentially continuous  action space.  $\mathcal{P}\left(s^{\prime} |s,a\right)$ denotes the transition  probability to a next state $s'$ when  action $a$ is taken at state $s$.
Finally, $r(s,a) \in \mathcal{R}$ is the reward evaluated at state action pair $(s,a)$ that reflects the immediate reward,  and $\gamma\in (0,1]$ is a discount factor.
An agent's policy $\pi(a|s)$ is a probability distribution that defines the probability of actions given a state $s$. 
Since RL tackles sequential decision problems, we define the state-action value function $Q^{\pi}(s,a)$ that reflects the long-term value of taking action $a$ in a certain state $s$:
\begin{equation}
Q^{\pi}(s, a) \triangleq \mathbb{E}\left[\sum_{t=0}^{\infty}\gamma^{t}r\left(s_{t},a_t\right) \Big| s_{0}=s, a_0=a\right], 
\label{eq:action_value}
\end{equation}
where the expectation is with respect to the transition dynamics $P$ and policy $\pi$.
The action value function is the expected cumulative rewards starting from $s$ and drawing actions according to $\pi$.
We also define the state-only value of being in state $s$ as $V^{\pi}$(s).
It is called the average value as it has the property that $V^{\pi}(s) = \mathbb{E}_{\pi}[Q^{\pi}(s,a)]$.
The action value function plays a key role in most if not all algorithms in this survey, since if we have a finite number of actions $\{a_i\}$ and know exactly their values $Q(s, \{a_i\})$ for every state, then the optimal policy can simply be obtained by choosing greedily with respect to the $Q$ function i.e. $\argmax_a Q(s,a)$.

In the following, we first clarify the differences pertaining to the MDP quintuple about $\gamma$ (learning horizon) in Section \ref{sec:bandits_mdp}, transition dynamics (model-free/model-based) in Section \ref{sec:free_based}.
We introduce value-based, policy-based, and actor-critic methods in Section \ref{sec:value_policy} where the majority of off-the-shelf algorithms are available in public libraries.
We discuss in Section \ref{sec:deep_rl} other related topics like how deep learning has empowered simple reinforcement learning methods to super-human level performance.
Section \ref{sec:miscellaneous} is devoted to other RL topics like multi-agent RL and the difference between online and offline learning.

\subsection{Bandits and Sequential Decision Making Problems }\label{sec:bandits_mdp}

In our MDP definition, we defined that the discount factor $\gamma \in (0, 1]$, which ruled out the possibility that $\gamma=0$.
In fact, $\gamma=0$ corresponds to another class of important problems known as the multi-armed bandits (MAB).
In contrast to RL problems that focus on sequential decision making problems that have evolving states, MABs are stateless in that it has only one unchanged state, and the goal is to choose the best action with optimal reward.
MABs have been widely adopted in SE tasks such as mutation operator selection for fuzzing~\cite{yu2024multiple} and automated program repair (APR)~\cite{hanna2025reinforcement}. This is primarily because these tasks naturally align with the bandit setting, where no explicit notion of state is required. The core insight of MABs involves dynamically balancing exploration (trying potentially useful but underused operators) and exploitation (favoring operators that have shown high reward). In fuzzing or APR, exploring rarely used mutation operators may reveal new bugs or viable patches, while exploiting previously successful ones helps generate high-quality variants. More importantly, the stateless nature of MABs makes them especially suitable for scenarios like APR, where modeling program states is highly nontrivial due to the complexity of real-world software. In contrast, stateful RL algorithms like Q-learning require well-defined state representations and transitions. By treating each mutation operator as an independent arm and estimating its utility solely based on past rewards, MABs bypass the need for complex state modeling, offering a lightweight yet effective alternative to full RL formulations.

\subsection{Model-free and Model-based Algorithms}\label{sec:free_based}

RL algorithms can be categorized into the model-free and model-based classes depending on the availability of \emph{model}, which dictates how the environment behaves.
In the MDP quintuple, this model is given by the transition dynamics $P$.
Typically, in an RL problem one needs to \emph{explore} the environment, which  refers to estimating the transition dynamics $P(s'|s,a)$ and rewards $r(s,a)$.
If we can precisely characterize transition dynamics, then the environment is no longer a black box, and the RL problem becomes a planning problem.
If the reward function is also known (for example, defined by the user), then the optimal policy can be computed without any interaction with the environment using dynamic programming \citep{Bellman2003} (assume small state-action spaces).
Even if the reward is not known, the demand for exploration could be drastically reduced since learning can be more targeted. 

Unfortunately, learning an accurate model is a challenging task itself, except for some special cases.
Moreover, the learned model typically can be applied only to a limited range of scenarios due to the specifics of each learning regime.
By contrast, model-free RL is more general: the agent estimates transition dynamics and rewards from the interaction samples it gathers from the environment.
As such, model-free RL algorithms can be applied to any environment to have unbiased estimation of $P$ and $R$ and to learn optimal policies based on the estimates. 
The downsides of model-free RL are that it could require a large number of samples before it can attain the optimal policy.
This makes it unsuitable for applications where data is costly or difficult to obtain. In the context of SE, samples are often expensive to collect, as they are typically manually curated, filtered, and verified from open-source repositories, public websites, or industrial systems~\cite{yang2022survey,DBLP:conf/icse/YouWLC25}. As a result, model-based approaches are rarely adopted in RL-for-SE research.
However, we observe an exception in the domain of Android UI testing, where model-based methods have been successfully employed. Specifically, one study~\cite{ran2023badge} utilizes a model-based approach evaluated on 21 widely-used industrial Android applications, most of which have over 100 million installations. In this domain, both developers and researchers actively maintain high-quality datasets and are generally willing to share them with the community. This openness provides a foundation for the application of data-intensive methods such as model-based RL.

\subsection{Value-based, Policy-based and Actor-Critic Methods }\label{sec:value_policy}

The goal of RL is to obtain an optimal policy $\pi$.
RL algorithms can be roughly categorized into two broad classes of methods depending on how they obtain the policy.
The first class is the value-based method.
It leverages the action value defined in Eq. (\ref{eq:action_value}) based on the following simple idea: if we have an optimal action value $Q(s,a)$ for every state-action pair, then the optimal policy can be naively extracted by greedily selecting the best action $\pi(a|s) = \argmax_{a} Q(s,a), \forall s,a$.
Therefore, the problem is translated to estimating an optimal action value function, denoted by $Q^*$.
Given a policy, 
value-based methods seek to estimate their value by leveraging the Bellman equations:
\begin{equation}
B^{\pi}Q (s,a) \triangleq {r} (s,a) +\gamma \sum_{s'\in\mathcal{S}}\mathcal{P}(s'|s,a) \sum_{a'\in\mathcal{A}}\pi(a'|s')Q(s', a').
\label{eq:bellman_q}
\end{equation}
where $B^{\pi}$ is the Bellman operator \citep{Bellman2003}.
Bellman operator has a desirable property: it has a fixed point $Q^{\pi}$, that is, repetitively applying the operator on any $Q$ converges to the fixed point.
At the fixed-point, the operator no longer changes the value.
This property holds for the optimal policy $Q^*$ as well:
\begin{equation}
Q^* (s,a) \triangleq {r} (s,a) +\gamma \sum_{s'\in\mathcal{S}}\mathcal{P}(s'|s,a) \max_{a'}Q(s', a').
\label{eq:bellman_q_fixed_point}
\end{equation}
\textbf{Value based methods} utilize this property to estimate an optimal value function.
One of the most wide-used algorithm of this kind is Q-learning \citep{watkins1992q}.
It provides an easy-to-implement algorithm: starting from an arbitrarily initialized value $Q$, repetitively applying Eq. (\ref{eq:bellman_q}) until convergence and then acting greedily with respect to the learned value.
Q-learning is an off-policy algorithm, meaning that it can reuse past experience to update its $Q$ function.

Many studies~\cite{koroglu2018qbe,huang2024crashtranslator,wang2024mrca} adopt Q-learning as their RL method of choice, primarily for two reasons.
First, software systems under test are often highly complex and dynamic. For example, GUI states and interaction sequences in mobile applications frequently change based on user actions, such as page transitions or widget updates~\cite{huang2024crashtranslator}.
Similarly, microservice systems have evolving architectures, with shifting service dependencies and traffic patterns~\cite{wang2024mrca}.
These properties make it difficult to predefine accurate environment models, such as state transition probabilities. Q-learning, being model-free, can learn directly from real-time interactions with the system, making it well-suited for black-box testing and scenarios with high uncertainty.
Second, Q-learning is more simple and efficient. In contrast to brute-force search or Monte Carlo methods, which typically require a large number of action executions and result in high computational cost, Q-learning is more lightweight and practical~\cite{koroglu2018qbe}.
This is particularly important in software engineering tasks like testing, where the action space is large but discrete, and exploration efficiency is critical.

Value-based methods have solid theoretical underpinnings and enjoy relatively simple implementation. 
However, its downside lies in its limited capability for continuous action spaces.
In Eq.(\ref{eq:bellman_q}), solving $\max_a Q(s,a)$ in a discrete action space is typically straightforward, but it becomes intractable in a continuous space.
Policy-based methods tackle the issue of discrete action space from the policy perspective directly.
Specifically, policy methods define the objective $J(\pi)$ as a scalar summarizing the overall quality of the policy $\pi$.  
It further assumes that the policy  $\pi_{\theta}$ is parametrized by a vector of parameters $\theta$.
The policy can then be optimized by adjusting its parameters to the high reward region utilizing its gradient information $\nabla_{\theta} J(\pi; \theta)$.

The \textbf{Policy Gradient Theorem} \citep{Sutton1999} is the cornerstone of policy gradient methods.
It states that the policy gradient can be computed by the following:
\begin{equation}
    \nabla_{\theta} J(\pi;\theta)= \expectation{s\sim\rho^{\pi}, \, a\sim\pi_{\theta}}{Q^{\pi}(s, a)\nabla_{\theta}\ln\pi_{\theta}(a|s)}.
    \label{eq:policy_gradient}
\end{equation}
where $\rho^{\pi}$ denotes the stationary distribution induced by policy $\pi$.
Because the expectation is with respect to $\rho^{\pi}$ and $\pi_{\theta}$, it is an on-policy algorithm in that the policy being updated is the one that generates samples.
Practically, the current estimate of policy is deployed to interact with the environment for a certain time steps to collect samples for update.
After the update, all samples need to be discarded and the process repeated.
One important variant of Eq. (\ref{eq:policy_gradient}) is REINFORCE \citep{Williams-reinforce} that replaces $Q^\pi$ and $\rho^{\pi}$ with Monte-Carlo estimates:
\begin{equation}
    \nabla_{\theta} J(\pi;\theta)  = \expectation{\pi}{ {\sum_{t=0}^T\nabla_{\theta}\ln\pi_{\theta}(a_t|s_t)  
    }
    \AdaBracket{\sum_{\tau=t}^T \gamma^\tau R_{\tau}}}.
    \label{eq:reinforce}
\end{equation}
Essentially, REINFORCE can be implemented by \emph{rolling out} agent trajectories and recording cumulative rewards along the path starting from $\tau$ until the terminal time $T$.
The outer expectation with respect to $\pi$ can also be approximated by averaging over multiple such trajectories.

Because there is no restriction on the policy other than evaluating its score function $\nabla_{\theta} \ln \pi_{\theta}$, the policy can be parametrized as a continuous distribution, e.g., a Gaussian.
This way, policy gradient methods can handle continuous action spaces.
The Gaussian distribution is the most popular and standard choice, though researchers have explored more flexible choices such as the Beta distribution \citep{Chou2017-BetaRL} or more generally $q$-exponential family \citep{Zhu2025-qExpPolicy} that can show better performance.

Many studies adopt policy gradient as their primary reinforcement learning approach. This preference can be attributed to two main reasons:
(1) Lack of ground truth data. In many SE tasks, high-quality annotated data are scarce. For example, in smart contract repair, there is a lack of labeled pairs of buggy and correctly repaired code~\cite{guo2024smart}. Similarly, in code adversarial attack tasks, ground-truth action sequences are often unavailable~\cite{yao2024carl}. In such scenarios, Policy Gradient methods are advantageous because they optimize policies through self-exploration, relying solely on reward signals from the environment without the need for large amounts of labeled data.
(2) Optimization over probability distributions. A notable advantage of PG is its ability to directly optimize the parameters of a policy represented as a probability distribution. This feature is particularly beneficial in SE tasks involving probabilistic context-free grammars (PCFGs), such as comment generation~\cite{shi2023machine} and input grammar inference~\cite{wu2019reinam}.
Since PCFGs rely on probability distributions, Policy Gradients align well with the task objectives by treating the policy (i.e., the rule distribution) as the direct target of optimization.

\textbf{Actor-Critic methods} combine the merits of both value-based and policy-based methods.
It can handle continuous action spaces with the actor ($\pi$) and estimate its value using the critic ($Q^{\pi}$).
Actor and critic cooperate in a way that the critic is updated to be more accurate in evaluating the actor, and the actor is updated towards higher quality decisions, and in turn increases the value of the critic.
In fact, this cooperation is alluded to in the policy gradient theorem Eq. (\ref{eq:policy_gradient}).
Practically, in addition to the policy,  the value function $Q_{\phi}$ is also parametrized, e.g., by a neural network.
Actor-critic methods have shown superior performance in general than value-based or policy-based methods alone, with successful examples including the proximal policy optimization (PPO) \citep{schulman2017proximal}, soft actor-critic (SAC) \citep{haarnoja-SAC2018}, to name a few.

\subsection{Deep Reinforcement Learning}\label{sec:deep_rl}

Section \ref{sec:value_policy} discussed policy-based and actor-critic methods that need to evaluate the gradients of the policy objective.
This can be done in a straightforward manner using an auto-differentiation package like PyTorch \citep{PyTorch}.
Even for value-based methods, deep networks can greatly improve the performance of simple algorithms like Q-learning \citep{mnih2013atari}.
In fact, deep neural networks have empowered RL to super-human level performance on various complex tasks such as video game playing or Go \citep{mnih2015human,Silver2016,Silver2017-GowithoutHumanKnowledge} due to their expressiveness in modeling the policy and value functions.
However, it is not straightforward to combine RL algorithms to deep learning per se.
We discuss several important considerations that play a key role in the success of those deep RL algorithms.

\textbf{Experience Replay. }
Off-policy algorithms can use experiences collected by other policies.
Therefore, they can also reuse past experience generated by past iterations of the policy.
Experience replay discusses storing the past experience and playing it to the agent in order to strengthen its learning efficiency \citep{Lin1992-experienceReplay}.
In the seminal Deep Q-Network (DQN) paper \citep{mnih2015human}, experience replay is introduced as a buffer $\mathcal{D}$ storing transitions $(s,a,r,s')$. 
When updating the network, these transitions are sampled from $\mathcal{D}$ in a random way to decorrelate the samples.
Some research has proposed more systematic ways to sample experiences so that samples with large errors are emphasized and presented to the agent more often to help accelerate learning \citep{Hindsight2017}.

\textbf{Target Network. }
Unlike the tabular setting, deep RL typically requires a long learning horizon that typically consists of millions of gradient steps.
The networks change at a fast pace and can result in unstable or collapsed learning.
To stabilize the training process, the DQN paper \citep{mnih2015human} introduces \emph{target network} that remains unchanged for a certain period of time.
For example, $Q_{\phi}$ can have a copy $Q_{\bar\phi}$ that is not updated directly.
After a period of timesteps, the network parameters $\phi$ are copied to $\bar{\phi}$ to catch up with learning progress.
Since $\bar{\phi}$ is relatively more stable, it can be used to produce target value estimates in the loss function, see below for an example.

\textbf{Deep $Q$ Function. }
Deep neural networks can significantly increase the expressiveness of $Q$ function to more accurately capture values of $(s,a)$.
To update the network, the fixed-point property of Bellman equations  Eq. (\ref{eq:bellman_q}) is used: since at the fixed point the left-hand side should equal the right-hand side, the difference between the two sides can be evaluated as a loss for minimization.
Practically, this leads to the following mean-square error:
\begin{equation}
    \mathcal{L}(\phi) = \expectation{\substack{(s,a,r,s')\sim\mathcal{D}\\a'\sim\pi(\cdot|s')}}{ 
 \AdaBracket{ Q_{\phi}(s,a) - r(s,a) - \gamma Q_{\bar\phi}(s',a')}^2 },
\end{equation}
where $\gamma Q_{\bar\phi}(s',a')$ is a single-point  approximation of the future quantity $ \gamma\sum_{s'\in\mathcal{S}}\mathcal{P}(s'|s,a) \sum_{a'\in\mathcal{A}}\pi(a'|s')Q(s', a')$.
Because the expectation with respect to the transition dynamics $\mathcal{P}$ and policy $\pi$ is approximated using a single sample, typically model-free methods require a large number of training steps (learning horizon) to converge.

Many studies adopt DQN algorithms as their primary reinforcement learning approach. This preference is mainly driven by two considerations:
(1) SE tasks often have discrete action spaces. Therefore, DQN becomes a natural choice.
When the number of candidate actions is relatively small, DQN can typically guarantee reasonable performance.
Its simple implementation and the off-policy nature render DQN a preferred first choice for various problems~\cite{zhang2023resplay}. 
(2) DQN has been verified to be effective in problems with a high-dimensional state space but a simple action space.
In playing Atari games \citep{mnih2015human}, the states are screenshots of video games but the actions are as simple as moving left or right.
This makes DQN a favored choice in SE tasks like the game load testing~\cite{tufano2022using}, where the input also consists of raw, high-dimensional game screenshots. 
In fog computing resource allocation tasks~\cite{lakhan2022efficient}, the state may include multidimensional attributes such as total system cost, computational capacities of fog nodes, and more. DQN is capable of effectively processing and learning from such complex state representations.

\subsection{Miscellaneous}\label{sec:miscellaneous}

\textbf{On-policy versus Off-policy.}
On-policy refers to learning from experiences generated by the learning policy itself, while off-policy methods can learn from arbitrary experiences. 
Within the deep reinforcement learning framework, the difference becomes more subtle: a policy is typically parametrized by a neural network and subject to constant updates.
Therefore, theoretically, the policies before and after a single update are different, and the samples generated by the previous policy become off-policy and cannot be reused for the next update.
In off-policy algorithms, experience replay that stores the transitions generated by past policies becomes a standard component: mini-batches of transitions are sampled randomly from the replay buffer for update.
In its extreme form, off-policy becomes offline learning \citep{Levine2020-offlineRL}  in that it learns completely from a dataset (also known as batch RL \citep{Lange2012-batchRL}).
Offline RL has become more and more popular and important in the real-world deployment of RL, such as in large language models.

\textbf{Reward Design.}
Reward design is of vital importance to RL problems, if not the most important.
Rewards define the agent's behavior and reflect the designer's belief of the \emph{desired behavior}.
One widely adopted strategy in reward design is punishment shaping, which involves assigning negative rewards (penalties) to discourage undesired behaviors during learning. 
This approach helps the RL agent avoid unsafe, inefficient, or suboptimal actions by explicitly signaling failure or violation of constraints.
For example, in concurrency testing tasks, Mukherjee et al.~\cite{mukherjee2020learning} assigned a large penalty (-1000) to the action of injecting a failure message, as this action could terminate the exploration process prematurely and impede learning.
Another common strategy is multi-level reward, where different layers or granularities of feedback are provided based on varying degrees of task completion or quality.  
Instead of issuing a binary success/failure signal, this approach assigns intermediate rewards that reflect partial progress or varying levels of correctness. For example, Doreste et al. ~\cite{doreste2024adversarial} trained an adversarial autonomous driving system (ADS) using a reward function that combined two components: one encouraging the agent to drive effectively in the given scenario, and the other guiding it toward a collision with the ego vehicle to fulfill the adversarial objective.

\textbf{Exploration Strategy.}
The exploration-exploitation trade-off is at the core of RL.
Namely, the agent has to decide when to stop exploring the environment and conclude that it has gathered enough information to exploit by acting greedily with respect to it.
When the state-action space is large, the agent needs significant exploration before it can have an accurate estimate of $Q(s,a)$ for the majority part of the space. 
The agent must carefully choose how much to continue exploring the space to accumulate more evidence and then act by $\argmax_{a}Q(s,a)$.
If it stops exploration too early, then the available options can be highly suboptimal, but if it spends too much time exploring, learning becomes slow or even impossible.

The $\epsilon$-greedy strategy is perhaps the simplest and most widely used exploration method.
It simply specifies that with small probability $\epsilon$, it picks a random action to explore, and with probability $1-\epsilon$ it acts greedily with respect to the current estimate.
This method can be implemented in a line of code by sampling a variable uniformly from $[0,1]$, then checking if it is smaller than $\epsilon$; if it is, randomly pick an action.
Despite its simplicity, $\epsilon$-greedy performs surprisingly well and can sometimes perform better than sophisticated exploration methods.

Curiosity-based (or intrinsically motivated) algorithms are a class of sophisticated methods typically dealing with sparse rewards. 
Sparse rewards are typical in environments that do not award intermediate steps but only upon reaching the goal.
For example, in the game of GO, it is unwise to assign rewards to intermediate steps since that would bias the agent into potentially globally suboptimal strategies.
Then from a reward perspective, the agent receives zero reward for every step until it wins or loses (which it can receive e.g., $+1$ or $-1$ reward). This poses a great challenge for learning.
Curiosity-based methods encourage the agent to explore novel or uncertain states as an auxiliary reward.

\section{Comparative Analysis}
\label{comparative}
In this section, we adopt the comparative analysis protocol proposed by \citet{Feldt_08} to systematically compare and highlight the novelty of our review relative to existing systematic studies. 

To identify relevant prior works, we used the following search string: ``Software Engineering'' AND (``systematic review'' OR ``mapping study'') AND (``deep learning'' OR ``machine learning'' OR ``reinforcement'') across major scholarly databases, including the ACM Digital Library, IEEE Xplore, and ScienceDirect. 
We excluded papers that (1) did not explicitly indicate their nature as systematic reviews in either the title or abstract, or (2) were not published in venues traditionally associated with the software engineering research community, or (3) did not specifically focus on a certain SE topic such as defect prediction, or (4) did not target the application of SE for learning-based systems.
The search yields a total of four systematic reviews~\citep{wang2022machine, yang2022survey, watson2022systematic, abo2023role} before 2025.
For each systematic review, we characterize it based on its research objectives, inclusion criteria, number of analyzed papers, and adopted methods of analysis.
Table~\ref{table:comparative} presents the summary of relevant literature reviews.
Below, we provide a comprehensive discussion of how our work differs from existing systematic reviews.

\textbf{Difference in Research Goals.} In the literature, three systematic reviews~\citep{wang2022machine, yang2022survey, watson2022systematic} target the adoption of deep learning (DL) or machine learning (ML) algorithms for general software engineering tasks.
Specifically, \citet{watson2022systematic} conducted a systematic literature review of 129 papers published between 2009 and 2019 to investigate the application of DL algorithms in SE research. 
Similarly, \citet{yang2022survey} reviewed 146 papers from 2006 to 2020, providing a comprehensive summary, classification, and analysis of DL applications across various SE tasks. 
Going beyond DL, \citet{wang2022machine} explored the broader use of ML and DL in SE by analyzing 1,428 ML/DL-related SE papers published between 2009 and 2020. 
Their focus was on understanding the complexity involved in applying ML/DL algorithms to SE problems and how this complexity affects the reproducibility and replicability of such studies.
Although these three studies included the keyword ``reinforcement'' in their search strategies, none of them conducted a focused or in-depth investigation into the unique characteristics, algorithms, and challenges specific to RL applications in software engineering. In contrast, our study systematically examines RL-for-SE literature, offering detailed insights into RL-specific contributions, such as targeted SE tasks, methodological characteristics, and domain coverage that were overlooked in prior broad-spectrum ML/DL reviews.
In recent years, RL has garnered increasing attention and shown promising potential across a variety of SE tasks.
Given this momentum, it is both timely and necessary to conduct a comprehensive literature review to systematically understand the current landscape of RL-for-SE research and to provide valuable guidance for researchers interested in exploring RL application in the SE domain.

\begin{table}[t]
\caption{Summary of relevant literature reviews. TSE (IEEE Transactions on Software Engineering); CSUR (ACM Computing Surveys); TOSEM (ACM Transactions on Software Engineering and Methodology); IST (Information and Software Technology).}
\label{table:comparative}
\begin{tabular}{llccc}
\toprule
 & DL for SE & Surveyed Papers & Surveyed Time & Publication Venue \\ \midrule
\citet{wang2022machine} &   ML/DL for SE           & 1,428               & 2009--2020             & TSE       \\
\citet{yang2022survey}  & DL for SE                & 146               & 2006--2020             & CSUR         \\
\citet{watson2022systematic}   & DL for SE                & 129               & 2009--2019             & TOSEM   \\
\citet{abo2023role}  &    RL for Software Testing  & 40               & 2012--2021 & IST   \\ \bottomrule
\textcolor{blue}{Ours} & \textcolor{blue}{RL for SE} & \textcolor{blue}{115} & \textcolor{blue}{2015--May 2025} &
\end{tabular}
\end{table}

\textbf{Latest Time Frame with a Broader and Comprehensive Perspective.}
In the existing literature, only one prior review study~\cite{abo2023role} specifically focuses on the application of RL in the SE domain, with its scope limited to software testing.
\citet{abo2023role} conducted a review of 40 papers published up to 2021 to examine the adoption of RL in software testing, including the types of software testing tasks where RL has been applied, as well as the commonly used RL algorithms and learning architectures.
Different from their work, our study extends the time frame to May 2025, capturing the most recent developments. 
Notably, our descriptive analysis (shown in Section~\ref{results:trend}) reveals a significant surge in RL-for-SE publications beginning in 2022.
As a result, our review includes a substantial number of studies, totaling 115 papers.
In addition, prior work did not account for the reputation or quality of publication venues. 
In contrast, our study specifically targets well-established and top-ranked venues within the SE domain, ensuring that the findings are both representative and reliable to better inform and guide future research directions.
Most importantly, our review broadens the scope from software testing to the broader field of software engineering, encompassing a diverse range of SE tasks. 
In particular, we examine the entire software development lifecycle, including software quality assurance, software development, and software maintenance, as detailed in Section~\ref{results:rq2}.
Moreover, to provide comprehensive insights, we not only report the basic trends and the RL algorithms applied to various SE tasks, but also identify the key challenges and future opportunities in RL-for-SE research.
\section{Methodology}
\label{methodology}
\begin{figure}
    \centering
    \includegraphics[width=.8\linewidth]{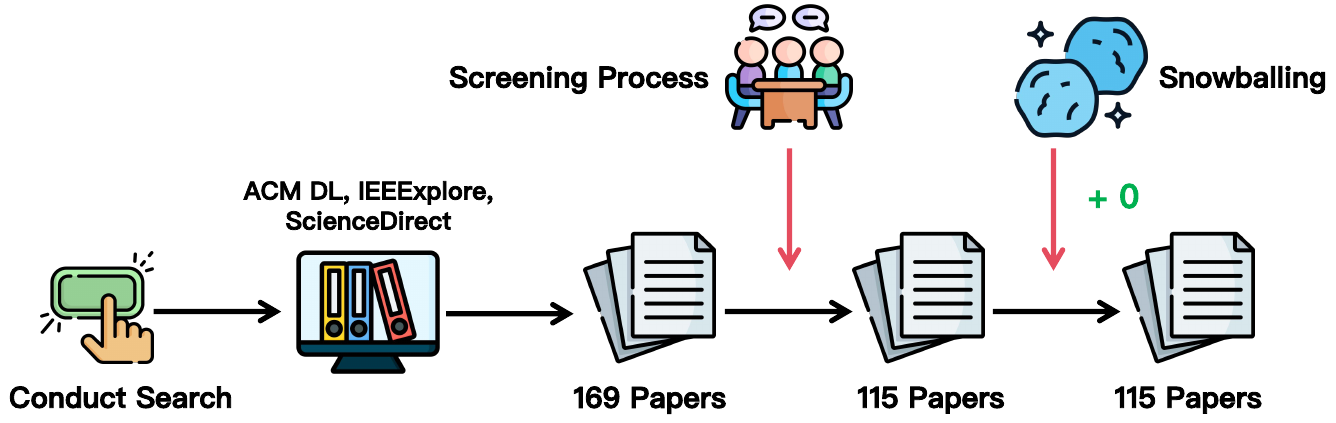}
    \caption{Overview of paper collection process}
    \label{fig:data_collection}
\end{figure}

Our systematic literature review follows the guidelines provided by~\citet{petersen2008systematic}. 
We begin by formulating the research questions to define the scope of the survey (Section~\ref{RQ}). Next, we carry out the search and screening processes to identify qualified studies (Sections~\ref{conductsearch} and~\ref{screen}). Finally, we perform keywording to extract relevant data items for each research question (Section~\ref{keywording}). Figure~\ref{fig:data_collection} shows the overview of the paper collection process.

\subsection{Research Questions}
\label{RQ}
The goal of this survey is to systematically examine the current landscape of reinforcement learning adoptions in software engineering by retrieving, categorizing, and analyzing relevant literature. 
To structure our review and ensure comprehensive coverage, we propose and address the following five research questions (RQs):

\begin{enumerate}
\item \textbf{RQ1: What are the trends in the primary studies on the use of RL in SE?} In this RQ, we conduct a demographic analysis to provide a comprehensive and visual overview of RL-related survey studies in SE, focusing on the distribution of publication years, venues, contributing authors, and types of research contributions.

\item \textbf{RQ2: What types of SE tasks have been facilitated by RL-based approaches?} This RQ explores the types of SE tasks and development phases that have been facilitated by RL, aiming to uncover insights into enhancing the applicability and generalizability of RL-driven studies.

\item \textbf{RQ3: What RL algorithms have been applied to support SE tasks?} This RQ presents a classification of RL algorithms employed in SE tasks and analyzes their popularity based on usage frequency across surveyed studies over time and across task types.

\item \textbf{RQ4: What are the model designs and evaluation practices in RL-for-SE tasks?} This RQ aims to gain deeper insights into RL-based approaches used in SE tasks by examining the construction of RL frameworks from four key aspects: evaluated data, model design and optimization strategies, evaluation metrics, and the replicability of the surveyed studies.

\item \textbf{RQ5: What challenges and opportunities emerge from the adoption of RL in SE?} This RQ outlines the research progress of RL-related studies, identifies unresolved challenges in existing work, and highlights potential opportunities to enhance the adoption of RL in SE, serving as a key contribution of this survey.

\end{enumerate}

\begin{table}[t]
    \centering
    \small
    \caption{Publication venues for conduct search}
    \label{tab:publication_venues}
    \begin{tabular}{llp{10cm}}
        \toprule
        \textbf{No.} & \textbf{Acronym} & \textbf{Full name} \\
        \midrule
        1  & ICSE   & ACM/IEEE International Conference on Software Engineering \\
        2  & ASE    & ACM/IEEE International Conference Automated Software Engineering \\
        3  & ESEC/FSE & ACM SIGSOFT Symposium on the Foundation of Software Engineering/European Software Engineering Conference \\
        4  & OOPSLA     & ACM SIGPLAN International Conference on Object-Oriented Programming Systems, Languages, and Applications \\
        5  & ISSTA  & International Symposium on Testing and Analysis Working Conference on Mining Software Repositories \\
        6  & ICSME  & IEEE International Conference on Software Maintenance and Evolution \\
        7  & ICPC   & IEEE International Conference on Program Comprehension \\
        8  & ESEM   & International Symposium on Empirical Software Engineering and Measurement \\
        9  & MSR    & IEEE Working Conference on Mining Software Repositories \\
        10 & SANER  & IEEE International Conference on Software Analysis, Evolution and Reengineering \\
        11 & ICST   & IEEE International Conference on Software Testing, Verification and Validation \\
        12 & ISSRE  & IEEE International Symposium on Software Reliability Engineering \\ \midrule
        13 & TSE    & IEEE Transactions on Software Engineering \\
        14 & TOSEM  & ACM Transactions on Software Engineering and Methodology \\
        15 & ESE    & Empirical Software Engineering \\
        16 & JSS    & Journal of Systems and Software \\
        17 & IST    & Information and Software Systems \\
        18 & ASEJ   & Automated Engineering \\
        19 & STVR   & Software Testing, Verification and Reliability \\
        20 & JSEP   & Journal of Software: Evolution and Process \\
        21 & SQJ    & Software Quality Journal \\
        22 & SCP    & Science of Computer Programming \\
        \bottomrule
    \end{tabular}
\end{table}

\subsection{Conduct Search}
\label{conductsearch}
This process aims to identify initial studies using search strings on scientific databases through relevant conference proceedings or journal publications as per \citet{petersen2008systematic}. 
To this end, three key parameters were considered to guide the search for initial studies:
\begin{itemize}[leftmargin=10pt]

\item \textbf{Databases}. Following prior survey studies~\cite{yang2022survey, wang2021can, wang2022machine} and to ensure the inclusion of high-quality research that reflects the state-of-the-art in the field, we selected 22 well-established and widely recognized venues in software engineering, comprising 10 journals and 12 conferences listed in Table~\ref{tab:publication_venues}. 
To minimize selection bias, we retrieved studies from a diverse set of digital libraries, including IEEE Xplore, ACM Digital Library, and ScienceDirect.

\item \textbf{Search String}. We first identified a set of RL-related keywords commonly used in SE publications involving RL algorithms. 
We then refined the search string iteratively by examining the titles and abstracts of a sample of relevant papers to ensure comprehensive and accurate coverage.
Subsequently, we applied logical OR operators to combine these terms, yielding the following search string: (\textit{
``reinforcement'' OR ``Q-learning'' or ``Q-Network'' or ``award''}).

\item \textbf{Time Interval}. We set the retrieval time interval from January 2015 to May 2025, covering a span of approximately ten years. 
Inspired by the influential works of \citet{lecun2015deep} and \citet{mnih2015human}, we selected 2015 as the starting point, as this year marked a significant surge in the popularity of reinforcement learning, driven by major breakthroughs and high-profile successes in AI research.
For instance, a milestone work by \citet{mnih2015human} combined deep neural networks with Q-learning, enabling RL to scale to high-dimensional inputs such as raw pixels from Atari games, and demonstrating that an RL agent could achieve human-level performance.

\end{itemize}
Following the above parameters, we performed the search in the aforementioned databases for titles, keywords, and abstracts.
Ultimately, after eliminating duplicate papers, we identified a total of 169 studies (98 conferences + 71 journals).

\subsection{Screening Process}
\label{screen}
The screening process involved applying predefined inclusion and exclusion criteria to eliminate unqualified studies, such as those not published as full papers. 
To do so, two authors of this paper conducted a manual review by reading each paper's abstract to assess its eligibility.
The following inclusion and exclusion criteria were used to guide the selection of relevant studies:

\ding{52} Studies must be written in English.

\ding{52} Studies must adopt RL algorithms to address SE problems.

\ding{52} The length of the paper must not be less than 8 pages.

\ding{56} Books, keynote records, non-published manuscripts, and literature reviews are dropped.

\ding{56} Studies whose research objective is not to test RL-based systems or RL algorithms.

\ding{56} If a conference paper has an extended journal version, the conference version is excluded.

As a result, 54 papers were excluded during this step, and 115 studies that met the criteria were retained for our survey.
The final step involved manual \textbf{snowballing} to identify additional relevant studies that may have been missed during the initial retrieval process. Specifically, we reviewed the reference lists of the selected papers to capture related works whose titles might not contain the predefined query terms.
The snowballing process did not yield any additional papers, indicating that no relevant studies were missed in the initial retrieval.

\subsection{Keywording and Data Extraction}
\label{keywording}

\begin{table}[t]
    \centering
    \caption{Data items collection for each research question}
    \label{tab:data-items}
    \begin{tabular}{lll}
    \toprule
    \textbf{RQs} & \textbf{Extracted data items} & \textbf{Description} \\
    \midrule
    RQ1 & Publication year/venue/affiliations & Basic information of each surveyed paper. \\
    RQ1 & Contribution & The main category of contribution in each surveyed paper. \\
    RQ2 & Research perspective & Research purpose and relevant entities of each surveyed paper. \\
    RQ2 & Research topic & The category of specific SE topics in each surveyed paper. \\
    RQ3 & Algorithm & The classification for RL algorithms used in each surveyed paper. \\
    RQ3 & Rationale & The rationale behind algorithm selection. \\
    RQ4 & Data source & Such as open-source data, industrial data, or collected data. \\
    RQ4 & Model design & Specific designs in RL frameworks (e.g., state and reward). \\
    RQ4 & Optimization technique & Specific optimization techniques used (e.g., exploration and training). \\
    RQ4 & Metric & Metrics are used to evaluate the RL model (i.e., effectiveness and efficiency). \\
    RQ4 & Replicability & Presence/absence of replication package and its completeness. \\
    RQ5 & Challenge & Unsolved challenges in the existing surveyed papers. \\
    RQ5 & Opportunity & Potential opportunities in future studies. \\
    \bottomrule
    \end{tabular}
\end{table}

After retrieving the selected papers, we perform keywording and extract essential data items to support answering the research questions posed in this study.
Table~\ref{tab:data-items} presents the detailed mapping between the extracted data items and the corresponding research questions. 
For example, in terms of RQ4, we focus on five data items, including data source (e.g., open-source data, industrial data, and collected data), model design (e.g., state and reward), optimization techniques (e.g., exploration and training), evaluation metrics, and replicability.
To ensure the quality of data extraction and minimize subjectivity, the process was conducted in two stages (except for quantitative analysis of publication trends and venues). 
In the first stage, two authors independently performed manual data extraction. In the second stage, a round-table discussion was held involving the first author to resolve disagreements and reach a consensus.
\section{RQ1: Trends of RL-for-SE studies}
\label{results:trend}

\begin{figure}[t]
    \centering
    \subfloat[Numbers of publications per year\label{fig:temp}]{
    \includegraphics[width=.49\textwidth]{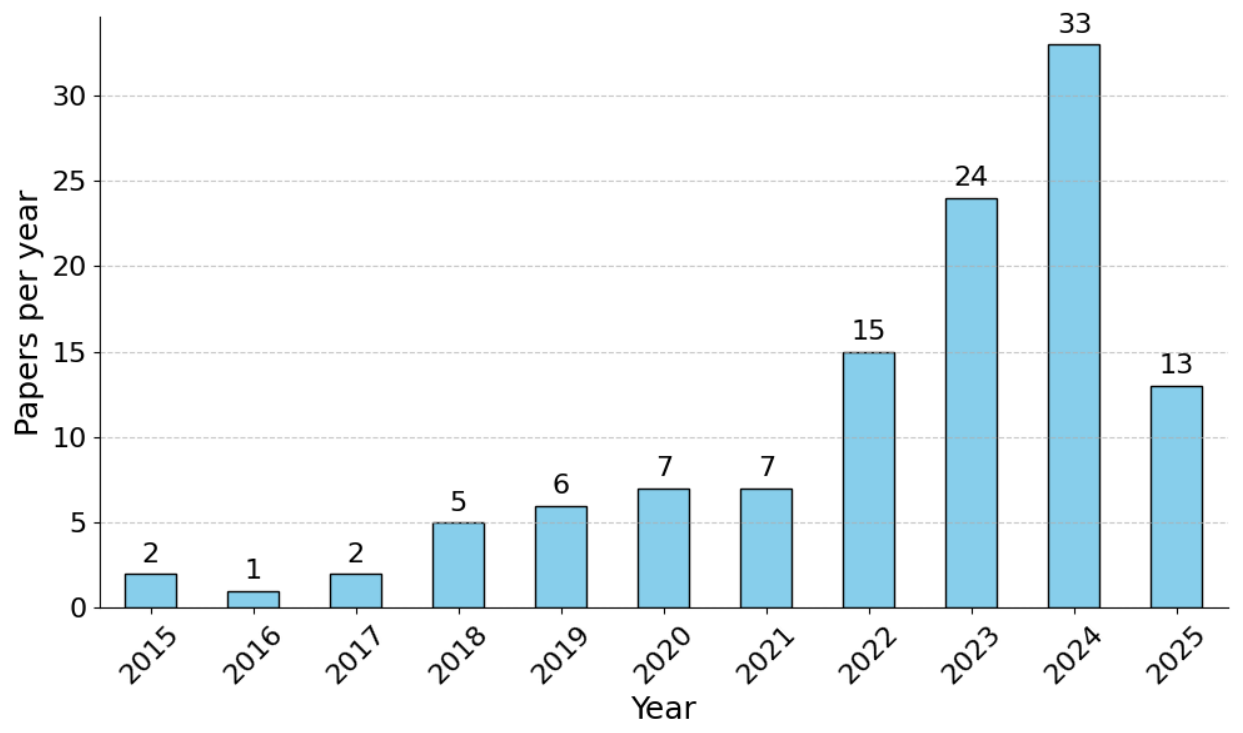}
    }
    \subfloat[ Cumulative number of publications per year\label{fig:mutation-order}]{
        \includegraphics[width=.49\textwidth]{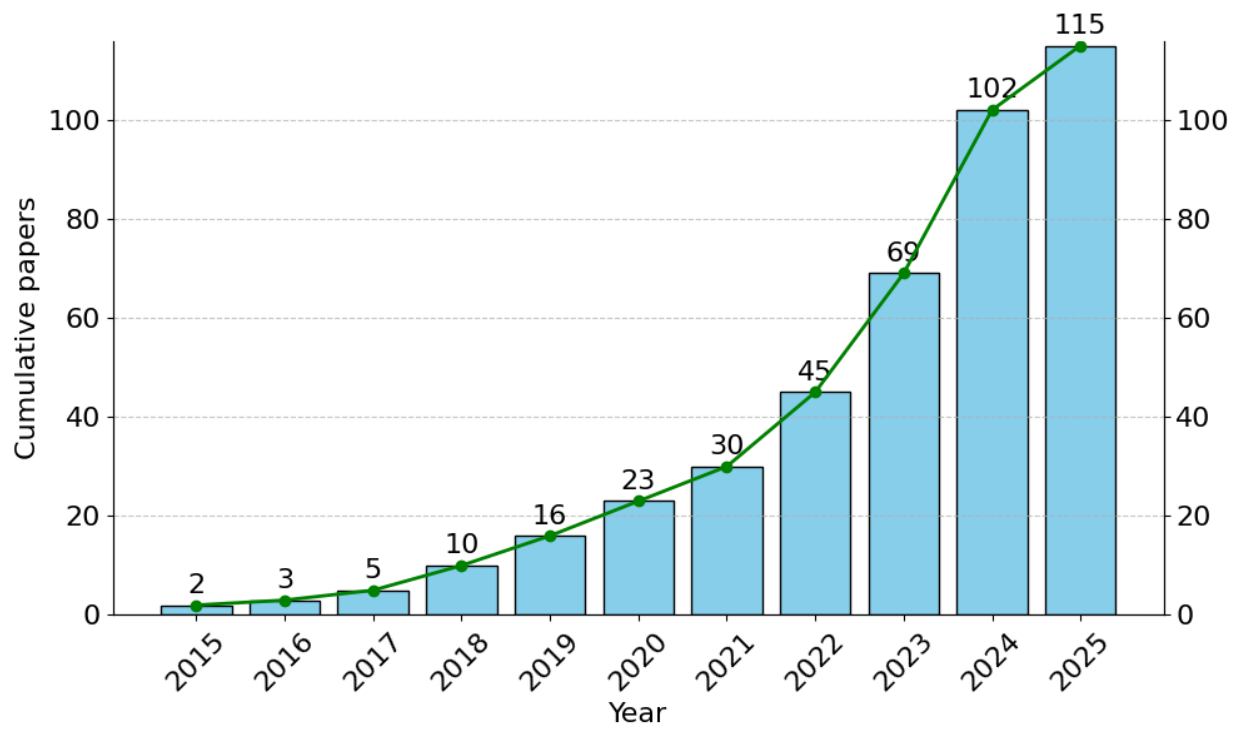}
    }
    \caption{Publication trends of surveyed RL-related studies from January 2015 to May 2025}
    \label{fig:trend}
\end{figure}

\begin{figure}[t]
    \centering
    \subfloat[Number of papers published in SE conferences\label{fig:conference}]{
    \includegraphics[width=.49\textwidth]{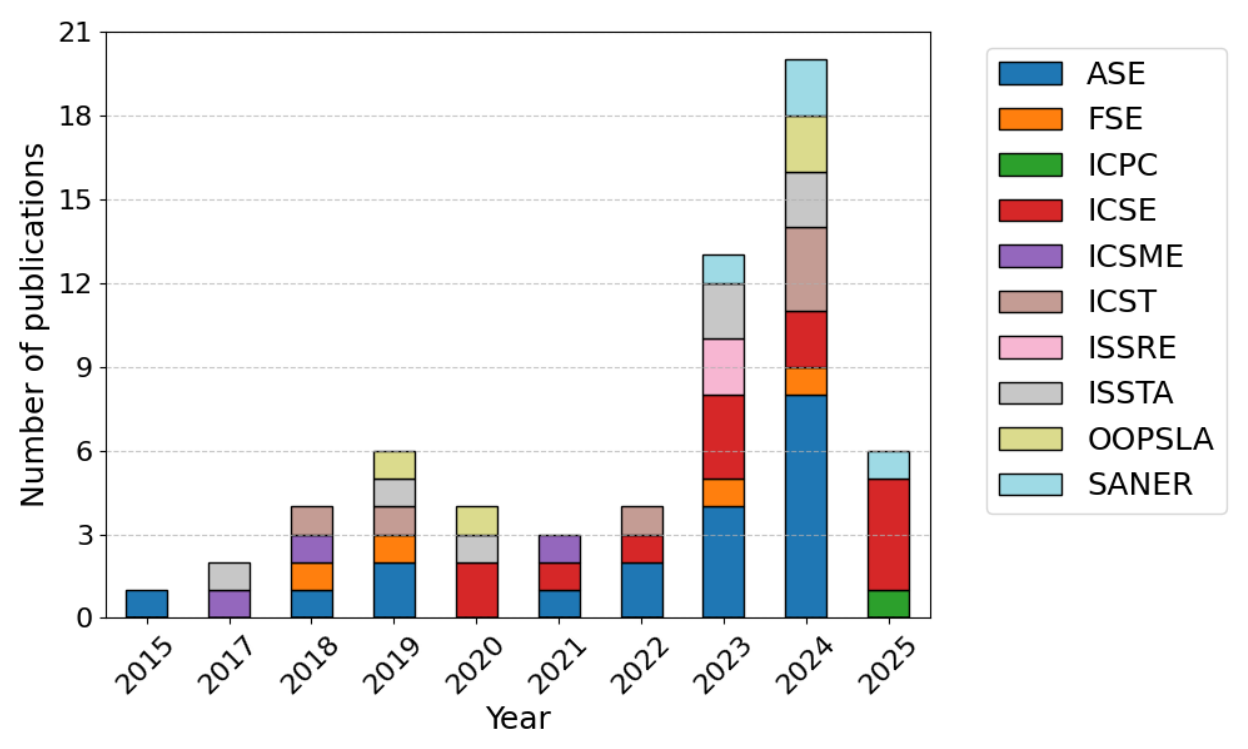}
    }
    \subfloat[Number of papers published in SE journals\label{fig:journal}]{
        \includegraphics[width=.49\textwidth]{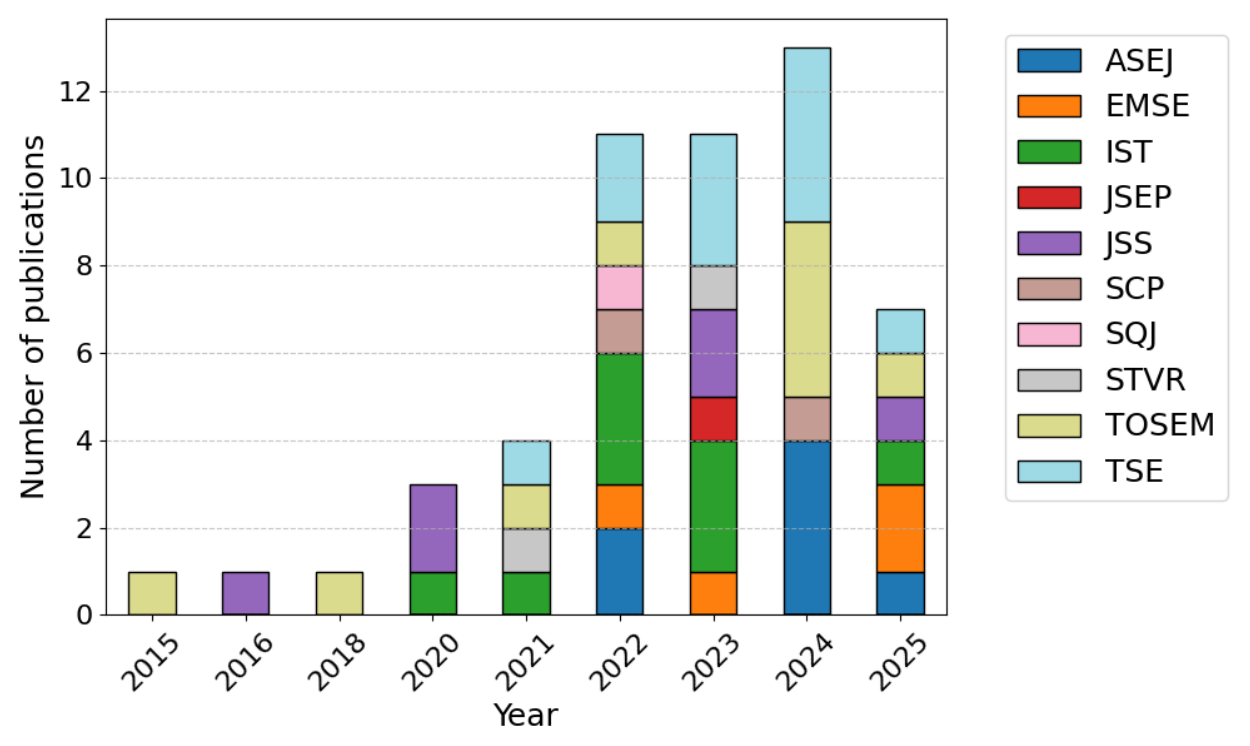}
    }\\
    \subfloat[Overall distribution across conferences and journals\label{fig:allvenue}]{
        \includegraphics[width=.4\textwidth]{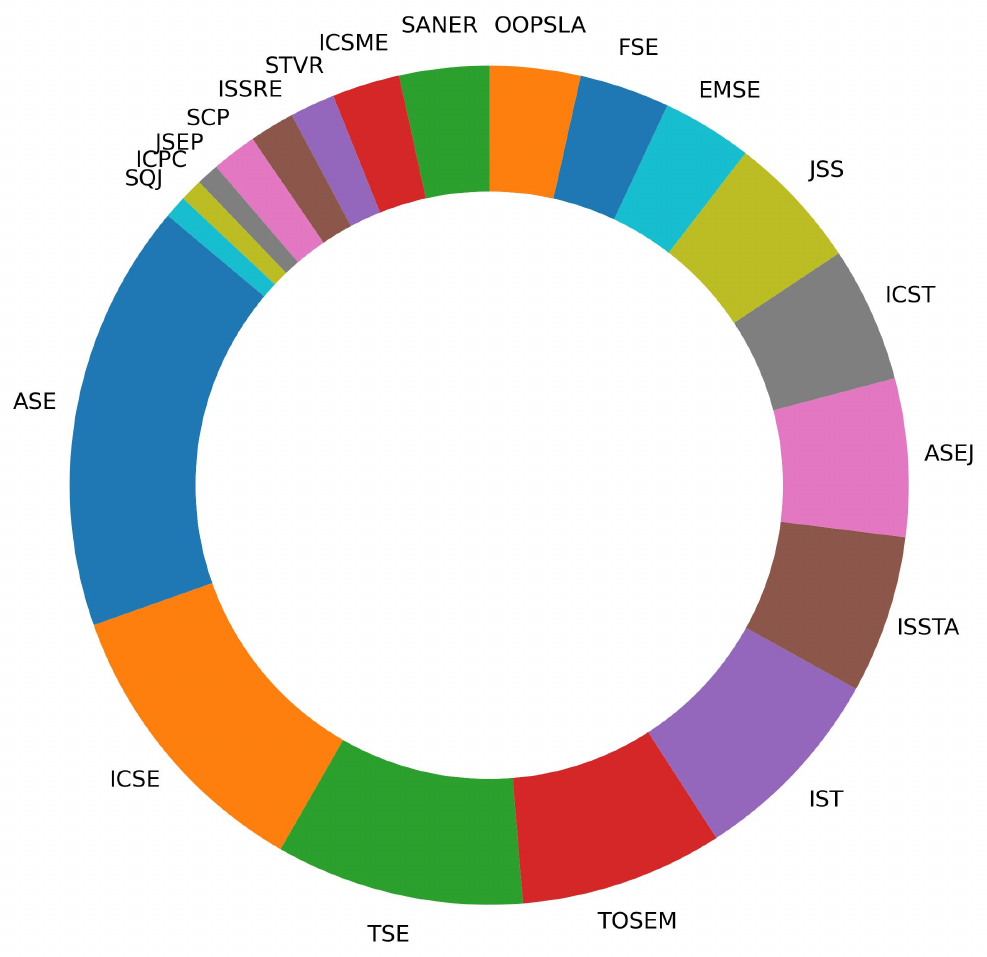}
    }
    \caption{Publication trends for SE conferences and journals from January 2015 to May 2025. Note that for 2025, due to conference schedules, only proceedings from ICSE, MSR, ICPC, and SANER are included.}
    \label{fig:venue}
\end{figure}

Figure~\ref{fig:trend} shows the publication trends of surveyed RL-related studies from January 2015 to May 2025, while Figure~\ref{fig:venue} shows the distribution of their published venues (i.e., journals and conferences).
Table~\ref{table:contribution} presents the distribution of contribution types.

\textbf{Publication trends in research on RL-for-SE.} As shown in Figure~\ref{fig:trend}, the number of RL-for-SE papers published per year exhibits a clear upward trend. 
Notably, even within the first quarter of 2025 before most conference proceedings have been released, six relevant papers have already been retrieved, underscoring growing research interest in the application of reinforcement learning to software engineering.
Specifically, the surveyed papers over the past decade can be broadly categorized into three phases: the initial stage (2015--2017), the growth stage (2018--2021), and the explosive stage (from 2022 onward). 
Remarkably, the number of publications peaked in 2024, with a total of 33 papers.

The cumulative plot in Figure~\ref{fig:venue} also illustrates the rapid growth of RL-for-SE publications over the past three years. Specifically, 85 new papers have been published since 2022, accounting for approximately 74\% of all surveyed studies.
As evidenced by longitudinal data, the application of RL in SE has shown a clear upward trajectory, reflecting growing research attention and adoption. 
This trend highlights the increasing recognition of RL as a promising approach for tackling a variety of SE challenges. Accordingly, a continued rise in RL-for-SE studies can be reasonably anticipated, as the research community seeks to further explore and exploit learning-based algorithms across diverse SE domains.

\textbf{Distribution of Surveyed Studies in Publication Venues.} 
First, we observe that the overall distribution between conferences and journals is relatively balanced, with 63 papers (55\%) published in conferences and 52 papers (45\%) in journals.
However, a comparison between Figure~\ref{fig:conference} and Figure~\ref{fig:journal} reveals that, since 2023, there has been a noticeable shift toward publishing RL-for-SE studies in conferences rather than journals.
For instance, in 2024, the number of RL-for-SE papers published in conferences reached 20, compared to 13 in journals.
This trend may reflect a preference for the faster publication cycles and timely dissemination offered by conferences, especially given the rising popularity of reinforcement learning as a trending topic within the SE research community.
Second, as shown in Figure~\ref{fig:allvenue}, the top three conferences with the highest number of RL-for-SE publications are ASE (19 papers), ICSE (14 papers), and ISSTA (7 papers). Notably, ASE has consistently featured at least one RL-related paper per year, except in 2017 and 2020, suggesting it is a favored venue for researchers in this area.
On the other hand, among journals, the top three venues are TSE (11 papers), IST (9 papers), and TOSEM (9 papers). 
These findings further indicate that RL-for-SE studies are frequently published in top-tier venues, particularly ASE, ICSE, TSE, and TOSEM, all of which are ranked A$\ast$ in the CORE conference and journal rankings.\footnote{\url{http://portal.core.edu.au/conf-ranks/}}

\textbf{Top Published Countries.} 
We further analyze the distribution of top first authors and their affiliated countries across the surveyed papers. 
The results reveal that RL-for-SE research spans 24 countries, reflecting broad international engagement with this emerging area. 
Notably, China leads in terms of publication volume, with 47 papers accounting for 41\% of the total, underscoring its prominent role in advancing the application of reinforcement learning within the software engineering domain.
Following China, the most frequently represented countries are the United States (16 papers), Canada (12 papers), and Italy (8 papers), highlighting their active contributions to RL-for-SE research.

\begin{table}[]
	\fontsize{9}{11}\selectfont
	\tabcolsep=0.2cm
\centering
\caption{Distribution of contribution types. Note that some studies fall into multiple categories, e.g., presenting both a novel technique or methodology and an accompanying empirical study.}
    \label{table:contribution}
\begin{tabular}{lp{6cm}cl}
\toprule
\textbf{Category}  &                          & \textbf{Count} &  \\ \midrule
New technique or methodology    & The paper proposed a novel approach, a.k.a. framework,
algorithm, and solution for applying RL to SE tasks.  &  106 & \mybar{0.88}   \\
Empirical study   & The study collected primary data and performed a quantitative and qualitative analysis on the data to explore interesting findings. &  8&\mybar{0.07}   \\
Case study & The study analyzed certain SE issues based on one or more specific
cases. &  4 & \mybar{.03} \\
Tool & The study implemented and published a new tool or a system targeting
SE issues. &  1 & \mybar{.01} \\ 
User study & The study conducted a survey to investigate the attitudes of different
people (e.g., developers, practitioners, users, etc) towards SE issues.  & 1 & \mybar{.01} \\ 
\bottomrule
\end{tabular}
\vspace{-.4cm}
\end{table}

\textbf{Distribution of Contribution Types.} Following the approach of the prior literature review~\cite{yang2022survey}, we manually examined the primary contribution of each surveyed study and categorized them into four types: new technique or methodology, tool, empirical study, and user study.
A single paper may be assigned to multiple contribution types; the total count exceeds the number of surveyed studies (115).
As shown in Table~\ref{table:contribution}, the majority of papers were categorized as proposing a new technique or methodology, accounting for approximately 88\% (106 papers). This indicates that most RL-for-SE studies focus on introducing novel approaches, such as frameworks, algorithms, or solutions.
This is followed by empirical studies, with eight papers identified that comprehensively evaluate a range of RL algorithms applied to specific SE tasks.
More specifically, our classification reveals that four papers~\cite{fan2024can, hanna2025reinforcement, giamattei2025reinforcement, yang2020systematic} fall into both the `new technique or methodology' and `empirical study' categories. 
These papers first conduct a pilot study to empirically evaluate the performance of various RL algorithms, and subsequently leverage the gained insights to propose novel RL-based techniques.
For the case study category, four papers were identified, two of which specifically focus on autonomous driving issues~\cite{doreste2024adversarial, humeniuk2024reinforcement}.
The least frequent categories are user study and tool, with only one paper~\cite{ferretti2016automatic} falling into both. 
This paper presents a system that employs Web intelligence to perform automatic adaptations on individual elements of a web page and includes a real-world evaluation involving elderly users.

This distribution of contribution types indicates a strong emphasis on proposing new techniques or methodologies in current RL-for-SE research. 
However, it also highlights a gap in empirical validations and user-centered evaluations. 
To foster a more holistic understanding and practical adoption of reinforcement learning in software engineering, future research is encouraged to diversify its focus.
In particular, expanding rigorous empirical studies and conducting real-world user evaluations would not only strengthen the external validity of proposed approaches but also provide valuable insights into their usability and effectiveness in practical settings.

\begin{tcolorbox}[colback=gray!5,colframe=awesome,title= RQ1 Summary]
\textcolor{BurntOrange}{\faStar}~The number of RL-for-SE papers published each year has shown a clear upward trend, particularly since 2022, with conferences emerging as the dominant venues for dissemination.\\ 
\textcolor{BurntOrange}{\faStar}~A significant majority of the surveyed studies (88\%) focus on proposing new techniques or methodologies that apply RL to SE tasks.
However, there is a relative scarcity of work aimed at empirically understanding the effectiveness of RL across different SE tasks or evaluating the practical applicability of these techniques through user studies.\\
\textcolor{BurntOrange}{\faStar}~In terms of geographic distribution, researchers from China lead the field (41\% of publications), followed by contributions from the United States and Canada.
\end{tcolorbox}

\section{RQ2: Categorization of Software Engineering Tasks in RL-for-SE Research}
\label{results:rq2}

\begin{figure}
    \centering
    \includegraphics[width=\linewidth]{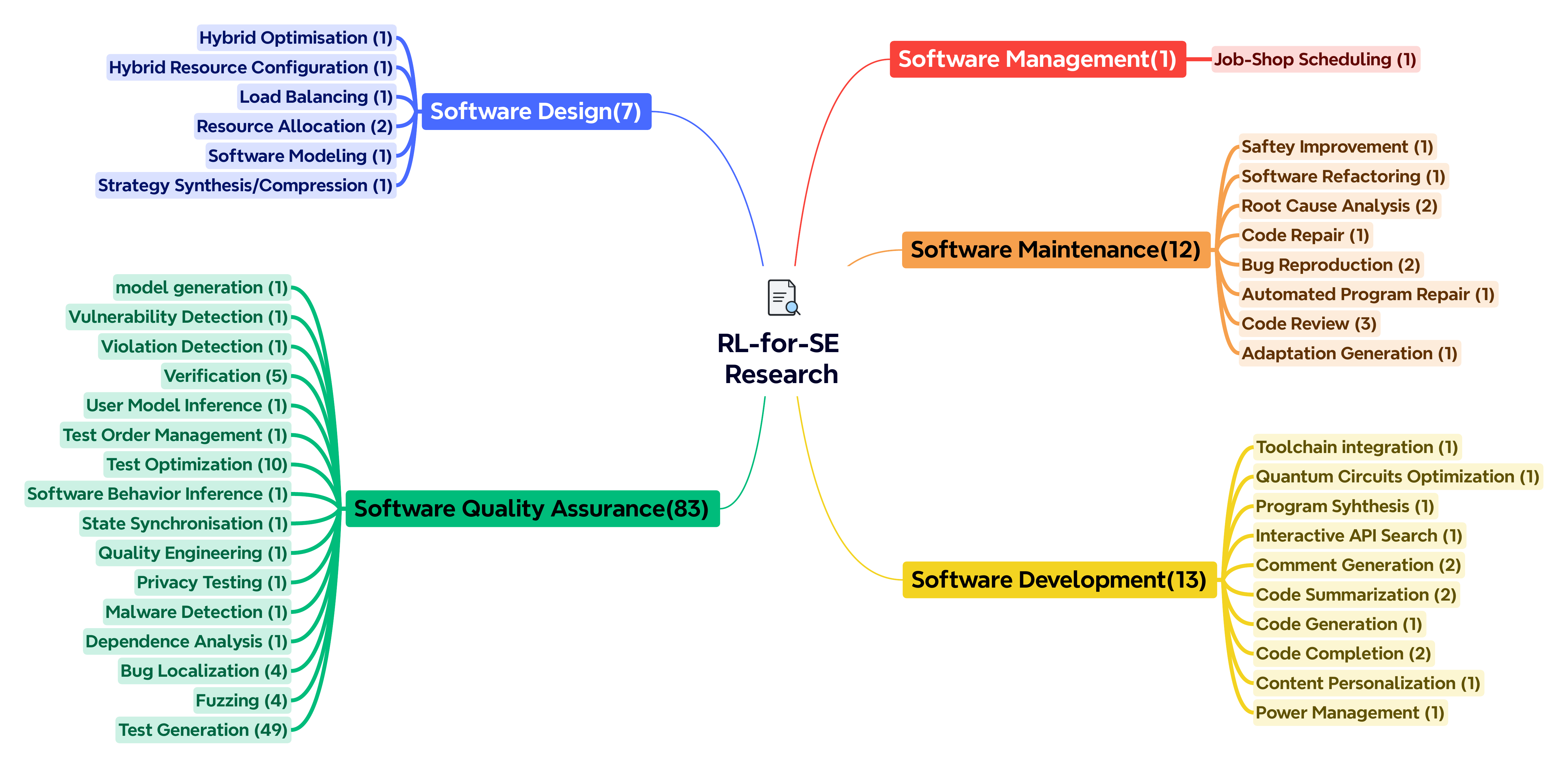}
    \caption{Distribution of SE activities and research topics utilizing RL}
    \label{fig:activity}
\end{figure}

In this RQ, guided by the Software Engineering Body of Knowledge~\cite{bourque2002guide}, we categorize a variety of SE tasks into six SE activities: (I) \textbf{Software requirements} (Focuses on eliciting, analyzing, specifying, and validating the needs and constraints of stakeholders to define what the software should accomplish), (II) \textbf{Software Design} (Involves defining the architecture, components, interfaces, and other characteristics to satisfy specified requirements), (III) \textbf{Software Development} (Encompasses the implementation and construction of software components through coding and integration), (IV) \textbf{Software Quality Assurance} (Ensures the software meets quality standards through activities like testing, verification, and validation), (V) \textbf{Software Maintenance} (Involves modifying and updating software after delivery to correct faults, improve performance, or adapt to a changed environment), and (VI) \textbf{Software Management} (Addresses planning, coordinating, measuring, monitoring, and controlling software projects and development processes).
In addition, we manually label the research topics under these six SE activities using the card-sorting method~\cite{spencer2009card} to ensure consistent and structured classification.
Note that a single surveyed paper may be associated with multiple SE activities or research topics if it addresses more than one aspect of the software development lifecycle.
Figure~\ref{fig:activity} presents a mind map illustrating the six SE activities alongside their associated research topics.

\textbf{Distribution of SE activities and research topics.} As shown in Figure~\ref{fig:activity}, Software Quality Assurance emerges as the most frequently targeted software engineering activity, with 83 papers categorized under this activity, accounting for 72\% of the surveyed studies.
The proportions of RL-for-SE papers focusing on Software Development, Software Maintenance, and Software Design are 11\% (13 papers), 10\% (12 papers), and 6\% (7 papers), respectively.
On the other hand, only one paper addresses Software Management, and none of the surveyed studies explicitly focus on Software Requirements. 
These findings suggest that the application of RL-for-SE is currently concentrated in the area of Software Quality Assurance, indicating a strong research emphasis on testing, verification, and fault detection tasks. 
This aligns with the inherently exploratory nature of RL, which fits well with tasks involving search-based strategies and uncertainty handling.
In contrast, Software Requirements and Software Management are significantly underexplored, with the former receiving no attention at all in the surveyed studies. 
This under-representation highlights important research gaps and opportunities. 
Overall, the current distribution calls for a more balanced exploration of SE activities in RL-for-SE research, encouraging the community to move beyond quality assurance and investigate how RL can address challenges across the entire software development life-cycle.

A deeper examination of the research topics under the categorized SE activities reveals that test generation stands out as the dominant focus within the Software Quality Assurance category. 
As shown in Figure~\ref{fig:activity}, 49 papers were labeled under this topic (accounting for 42.6\%), highlighting the strong alignment between RL and the task of automatically generating effective test cases. 
The next most frequent research topics are test optimization, with 10 papers, and verification, with 5 papers.
For another three SE activities (i.e., Software Development, Software Maintenance, Software Design), no dominant research topics were identified, as each contains only one or two related studies.

\begin{table}[]
	\fontsize{9}{11}\selectfont
	\tabcolsep=0.2cm
\centering
\caption{Distribution of SE task types}
    \label{table:tasktype}
\begin{tabular}{lp{8cm}cl}
\toprule
\textbf{Task Type}  &                          & \textbf{Count} &  \\ \midrule
Generation    & Automatically producing software artifacts based on specific inputs or contexts, such as test case generation and code completion.  &  85 & \mybar{0.74}   \\
Ranking   & Ordering a list of software items based on specific criteria, such as test optimization and verification &  21 &\mybar{0.18}   \\
Classification & Assigning predefined labels or categories to software-related entities, such as vulnerability detection and outcome prediction. &  5 & \mybar{.04} \\
Regression & Predicting continuous numeric values based on software data inputs, such as resource allocation and dependence analysis. &  6 & \mybar{.04} \\ 
\bottomrule
\end{tabular}
\end{table}

\textbf{Distribution of SE Task Types.} We additionally categorize the SE tasks of surveyed papers into the following four types: \textbf{Generation} (automatically producing software artifacts based on specific inputs or contexts), \textbf{Classification} (assigning predefined labels or categories to software-related entities), \textbf{Ranking} (ordering a list of software items based on specific criteria), and \textbf{Regression} (predicting continuous numeric values based on software data inputs).
Table~\ref{table:tasktype} presents the distribution of SE task types for surveyed RL-for-SE papers. 
Among the identified task types, generation is the most prevalent, with 85 papers (74\%) addressing it, followed by ranking tasks, which are covered in 21 papers (18\%).
In contrast, classification and regression tasks are comparatively underexplored, with only 5 and 6 papers identified, respectively.
These findings suggest that the existing RL-for-SE studies primarily look into the application of RL in the generation tasks.
These findings indicate that current RL-for-SE research is predominantly concentrated on generation tasks, reflecting a strong interest in leveraging reinforcement learning to produce outputs such as code, tests, or documentation. 

\smallskip
\noindent
Below, we present representative studies across the frequent research topics within the five categorized SE activities, illustrating how RL has been applied to address key challenges in each area.

\textbf{Software Quality Assurance - Test Generation.} This topic has been widely studied over the last decade, which aims to automatically generate test cases that can validate the correctness, robustness, or performance of software systems.
RL-based test generation has been explored in various domains, including fundamental software systems, domain-specific applications, and general-purpose software.
For \textit{complex fundamental systems}, \citet{chen2023compiler} proposed MCS, an RL-guided compiler testing approach that employs multi-agent reinforcement learning and memoized search to generate effective test configurations during on-the-fly compiler validation.
\citet{zhang2021figcps} proposed FIGCPS, a failure-inducing input generation approach that leverages deep reinforcement learning without requiring prior knowledge of the cyber-physical system (CPS) under test or access to historical logs, resources that are often difficult to obtain in practice.
\citet{le2024exploration} reported an industrial experience in testing a real-world R\&D flight control system by employing a heuristic-based testing method. 
Their approach integrates evolutionary strategies with intrinsic motivation to systematically generate a diverse set of test cases, each designed to uncover distinct potential failure scenarios within the flight control software. 

For \textit{domain-specific systems}, \citet{lu2022learning} introduced DeepCollision, which uses Deep Q-Learning to generate environment configurations likely to trigger collisions in autonomous vehicles. It models collision probability as a reward to guide scenario generation. 
\citet{humeniuk2024reinforcement} proposed RIGAA, combining RL with evolutionary search for testing autonomous systems. 
RIGAA trains an RL agent to guide initial population generation, improving failure discovery in robotics and vehicle lane-keeping systems.
To address the limitation of existing RL approaches that implement centralized controllers for the environment, leading to unrealistic or invalid behaviors of Non-Playable Characters, \citet{doreste2024adversarial} proposed an adversarial reinforcement learning approach for testing autonomous driving systems (ADSs). Their method focuses on training adversarial ADS agents using a custom reward function that balances driving quality with the likelihood of collisions, thereby avoiding overly aggressive or purely adversarial behaviors.

For \textit{general-purpose software}, \citet{pan2020reinforcement} presented Q-Testing, a curiosity-driven RL framework for Android testing that uses neural network-based state comparison to explore diverse app functionalities. \citet{reddy2020quickly} proposed RLCheck, an RL-based property testing method that formulates input generation as a diversifying guidance problem and uses on-policy table-based learning to adapt decisions during test time. 
\citet{nouwou2023comparison} conducted an empirical study on applying different RL algorithms to testing tasks in CI and interactive systems, showing the importance of algorithm-environment matching.
To address the limitation of existing REST API testing tools that often evaluate individual test elements in isolation, \citet{kim2024multi} proposed AutoRestTest, the first black-box testing framework that employs a dependency-aware multi-agent approach. 
AutoRestTest integrates Multi-Agent Reinforcement Learning (MARL) with a Semantic Property Dependency Graph (SPDG) and leverages Large Language Models (LLMs) to enhance the higher coverage and more effective fault detection.

\textbf{Software Quality Assurance - Test Optimization.} This research topic aims to improve the efficiency and effectiveness of software testing by minimizing testing cost while maintaining or enhancing fault detection capabilities.
The most prevalent RL-for-SE studies in this context focus on test case prioritization.
For instance, \citet{emam2015test} proposed a fault-based test case prioritization technique within model-based testing (MBT) frameworks. 
Their approach leverages a combination of reinforcement learning (Q-learning) and hidden Markov models (HMMs) to effectively identify and prioritize fault-revealing test cases, particularly in the context of regression testing.
To minimize the round-trip
time between code commits and developer feedback on failed test
cases in continuous integration (CI), \citet{spieker2017reinforcement} introduced
RETECS, which uses reinforcement learning (the policy gradient algorithm) to select
and prioritize test cases according to their duration, previous last execution, and failure history. 
\citet{yang2020systematic} performed an empirical study to systematically investigate the effect of the reward function and reward strategy in CI testing.
\citet{qian2025reinforcement} proposed a method TCP-KDRL that integrates LLEed (Linear Embedding) K-means clustering and dynamic priority factors into RL to mitigate the influence of adverse environments.

\textbf{Software Quality Assurance - Bug Localization.} It is the task of automatically identifying the most relevant locations in source code that are responsible for a reported software defect. 
It typically involves analyzing information such as bug reports, execution traces, or program spectra to rank or pinpoint suspicious code entities (e.g., files, classes, or methods). 
To facilitate compiler debugging, \citet{chen2020enhanced} introduced the first
reinforcement compiler bug isolation approach via structural mutation, called RecBi.
For a given compiler bug accompanied by a failing test program, RecBi enhances traditional local mutation operators to generate syntactically similar test programs. 
It further integrates reinforcement learning, specifically the Advantage Actor-Critic (A2C) algorithm, to leverage both historical and predictive knowledge in guiding mutation strategies. 
To address the limitations of similarity-based machine-learning
techniques for bug localization, \citet{chakraborty2024rlocator} proposed RLocator using a Markov Decision Process (MDP) to optimize the evaluation measures directly. 
The evaluation results on 8,316 bug reports confirm the promising role of RL in bug Localization.


\textbf{Software Quality Assurance - Fuzzing.} Fuzzing is an automated software testing technique that involves feeding a program with large volumes of invalid, unexpected, or randomly generated inputs to uncover bugs, crashes, or security vulnerabilities. 
\citet{wu2019reinam} proposed REINAM, a reinforcement learning-based approach for synthesizing probabilistic context-free grammars (PCFGs) for program inputs without relying on any seed inputs. 
Specifically, REINAM iteratively selects and applies generalization operators to the current PCFG, updates the production probabilities using a policy gradient reinforcement learning algorithm, and retains only those rules that are both general and accurate.
\citet{su2022effectively} introduced a novel vulnerability-guided fuzzer, Reinforcement Learning Fuzzer (RLF), which leverages the Deep Recurrent Q-Network (DRQN) algorithm to generate critical transaction sequences for effectively detecting vulnerabilities in smart contracts.
\citet{eom2024fuzzing} presented a fuzzing technique CovRL (Coverage-guided reinforcement learning) for JavaScript interpreters that combines LLMs with RL from coverage feedback.
Specifically, CovRL employs Term Frequency-Inverse Document
Frequency (TF-IDF) to weight coverage data, establishing a
feedback-driven reward system via the Proximal Policy Optimization (PPO) algorithm.

\textbf{Software Quality Assurance - Verification.} This research topic aims to ensure that software correctly implements specified requirements before it is released or deployed.
For example, \citet{chen2019relational} proposed a relational verification algorithm that incorporates RL to overcome the challenge of limited labeled training data for successful proof strategies. 
Specifically, their approach employs RL (the policy gradient algorithm) in an offline training phase to learn effective reduction strategies from a corpus of verification problems. 
In the online phase, the verifier applies the learned knowledge to efficiently solve new verification tasks.
\citet{liu2022learning} introduced an innovative approach for automatically learning contract invariants that support arithmetic safety verification, leveraging deep reinforcement learning. 
Their method features an offline training phase where the verifier iteratively learns a policy for generating useful contract invariants based on its own verification attempts.
Recently, to address the challenge of inefficient and undirected search in formal verification, \citet{sanchez2024qedcartographer} proposed QEDCartographer, an automated proof-synthesis tool that integrates supervised and reinforcement learning to more effectively navigate the proof space. By leveraging the branching structure of proofs, QEDCartographer enables reward-free exploration, thereby mitigating the sparse reward issue commonly faced in formal verification tasks.

\textbf{Software Development - Code Completion.} 
Code completion aims to improve programming productivity by automatically predicting and suggesting the next possible code elements based on the current development context.
To relieve inherent exposure bias of these models can cause errors to accumulate early in the sequence completion, \citet{li2024ircoco} proposed IRCoCo, a deep reinforcement learning (DRL)-based fine-tuning framework specifically tailored for code completion. 
IRCoCo introduces immediate reward feedback to detect and adapt to dynamic context changes resulting from continuous code edits. 
Experimental results show that fine-tuning pre-trained LLMs with IRCoCo significantly enhances code completion performance, outperforming both supervised fine-tuning (SFT) and other DRL-based baseline methods.
\citet{wang2024rlcoder} investigated repository-level code completion and proposed RLCoder, a reinforcement learning framework based on a weighted perplexity (PPL) mechanism. RLCoder enables the retriever to learn to identify useful content for code completion without requiring labeled data. Specifically, it iteratively evaluates the usefulness of retrieved content by measuring the perplexity of the target code when augmented with that content as additional context and uses this feedback to update the retriever's parameters.

\textbf{Software Development - Code Summarization.} This SE topic focuses on automatically generating concise, human-readable natural language descriptions of code snippets, enabling developers to quickly grasp the purpose and behavior of the code without examining it in full detail.
\citet{wan2018improving} integrated both the abstract syntax tree (AST) structure and the sequential content of code snippets into a deep reinforcement learning framework using an actor-critic architecture. 
Specifically, the actor network predicts the next word based on the current state with associated confidence, while the critic network estimates the reward values of all possible continuations of the current state, providing global guidance to enhance exploration during code summarization.
To enhance inferior and inconsistent comment generation accuracy, \citet{wang2020reinforcement} proposed a code summarization approach that employs a hierarchical attention network to integrate multiple code representations, including type-augmented ASTs, program control flows, and plain code sequences. These heterogeneous features are incorporated into a deep reinforcement learning framework, specifically an actor-critic network, to guide the generation of high-quality natural language comments describing code functionality.

\textbf{Software Development - Comment Generation.}
It focuses on automatically producing meaningful and natural language comments for code snippets. 
Unlike code summarization, which aims for concise descriptions, comment generation can produce more detailed and context-aware explanations tailored to the needs of documentation or onboarding new developers.
\citet{shi2023machine} proposed SolcTrans, an automated translation approach for Solidity smart contracts that generates fine-grained in-line natural language comments to enhance understanding, learning, and operation of smart contracts. 
SolcTrans utilizes the AST and formal grammar of the source code and employs reinforcement learning to train a Probabilistic Context-Free Grammar (PCFG)-based syntax synthesizer, which is responsible for producing coherent and comprehensible English sentences as code comments.
To address the gap in block comment generation, \citet{huang2020towards} proposed RL-BlockCom, a reinforcement learning-based approach that automatically generates block comments for code snippets using collected comment-code pairs. 
The method first transforms code into token sequences via statement-based traversal of its AST. 
Then, it employs a composite learning model that integrates an encoder-decoder architecture with an actor-critic reinforcement learning algorithm to generate informative and coherent block-level comments.

\textbf{Software Maintenance - Root Cause Analysis.}
This topic refers to the process of identifying the fundamental reasons why defects or failures occur in software systems, with the goal of preventing recurrence.
The existing work both focuses on microservice systems.
\citet{ding2023tracediag} proposed TraceDiag, an end-to-end root cause analysis (RCA) framework tailored for large-scale microservice systems. 
To tackle the complexity of service dependencies, TraceDiag employs reinforcement learning to learn an effective pruning policy that automatically filters out irrelevant components in the service dependency graph. This guided pruning significantly enhances RCA efficiency by narrowing down the analysis to only the most relevant services.
\citet{wang2024mrca} introduced MRCA, a metric-level root cause analysis (RCA) framework that leverages multi-modal data to enhance RCA effectiveness across diverse anomaly scenarios. MRCA begins by analyzing traces and logs to generate a ranked list of abnormal services based on reconstruction probabilities. 
It then constructs causal graphs from services with high anomaly likelihoods to uncover the temporal order of anomalous metrics across services. 
To avoid unnecessary graph expansion and improve efficiency, MRCA incorporates a reinforcement learning-based reward mechanism (i.e., Q-learning). This enables precise pruning of the ranked list and supports fine-grained identification of metric-level root causes.

\textbf{Software Maintenance - Bug Reproduction.}
This topic focuses on automatically re-executing the conditions that lead to a reported software bug, which is essential for effective diagnosis, debugging, and validation of fixes.
\citet{zhang2023automatically} proposed a fully automated approach for reproducing crashes from Android bug reports, addressing challenges related to limited natural language descriptions and incomplete or imprecise reproduction steps. 
The approach combines natural language processing to better interpret bug reports and reinforcement learning (i.e., Q-learning) to guide the search toward effective reproduction sequences.
\citet{huang2024crashtranslator} proposed CrashTranslator, an approach for automatically reproducing mobile application crashes directly from stack traces. 
It leverages a pre-trained Large Language Model to predict exploration steps likely to trigger the crash and incorporates a reinforcement learning-based technique to mitigate prediction inaccuracies and guide the search process more effectively.

\textbf{Software Maintenance - Code Review.} This topic focuses on assisting or automating various aspects of the process of systematically examining source code changes to identify bugs, enhance code quality, and ensure adherence to coding standards before integration. 
Typical tasks include change summarization, review outcome prediction, and review comment generation.
\citet{joshi2024comparative} explored factors that influence pull request (PR) outcomes and employs several RL formalizations (A2C, DQN, and PPO algorithms), modeled as Markov Decision Processes, for PR outcome prediction. 
To reduce the effort required by developers to write high-quality PR descriptions, \citet{liu2019automatic} proposed a novel encoder-decoder model that automates this task. 
To address the challenge of out-of-vocabulary words, their approach incorporates a pointer-generator mechanism that enables the model to copy words directly from the source code changes. 
Furthermore, they employ a reinforcement learning algorithm, Self-Critical Sequence Training (SCST), along with a specially designed loss function to directly optimize for the ROUGE metric, enhancing the quality and relevance of the generated summaries.
\citet{nashaat2024towards} proposed CodeMentor, a framework that facilitates few-shot learning to adapt large language models (LLMs) using organization-specific data for code review tasks such as code refinement and review comment generation. The framework constructs training datasets through heuristic rules and weak supervision techniques, enabling effective fine-tuning of LLMs with limited labeled data. To further incorporate domain expertise and improve review quality, CodeMentor integrates reinforcement learning with human feedback (RLHF), aligning the model's outputs with expert expectations.

\textbf{Software Design - Resource Allocation.}
This topic refers to the task of efficiently assigning limited computational or human resources to various components, modules, or processes within a software system, with the goal of optimizing system performance, maintainability, or overall cost-efficiency.
To address the challenge of resource allocation for application workloads in distributed and dynamic fog networks, \citet{lakhan2022efficient} investigated the resource allocation problem in Software-Defined Networking (SDN)-enabled fog environments. 
The authors proposed a novel container-based architecture incorporating heterogeneous fog nodes. Building upon this architecture, they introduced a Deep Q-Network (DQN)-based resource allocation strategy, which integrates key components such as a mobility controller, resource discovery, task scheduling, and task migration to effectively manage resources under dynamic network conditions.
\citet{tan2024energy} proposed a Multi-Domain Virtual Network Embedding (MD-VNE) algorithm leveraging deep reinforcement learning, specifically a single-step vanilla policy gradient method, to enable energy-efficient resource allocation for IoT data generation, which significantly improved the long-term revenue, long-term resource utilization, and allocation success rate when compared with existing methods.

\begin{tcolorbox}[colback=gray!5,colframe=awesome,title= RQ2 Summary]
\textcolor{BurntOrange}{\faStar}~Our findings show that RL-for-SE research is heavily concentrated in Software Quality Assurance (72\%), particularly in test generation, followed by Software Development (11\%), Software Maintenance (10\%), and Software Design (6\%). 
In contrast, areas like Software Requirements and Software Management are largely unexplored.\\
\textcolor{BurntOrange}{\faStar}~Task-wise, most studies focus on generation (74\%), such as producing tests or code, while ranking (18\%), classification (4\%), and regression (4\%) remain underrepresented.
\end{tcolorbox}

\begin{table}[t]
\caption{Summary of applied RL algorithms in surveyed papers}
\label{tab:tech}
\begin{adjustbox}{max width=1.0\textwidth, center}
\begin{tabular}{llp{4cm}p{6.5cm}p{2.5cm}r}
\toprule
        &    & Type                & Algorithms                                              & Related Papers & Total \\ \midrule
& \textbf{Model-free}  & Value-based         & Q-learning                                              & \cite{carino2015dynamically, koroglu2018qbe, koo2019pyse, pan2020reinforcement, mukherjee2020learning, zheng2021automatic, turker2021efficient, ferdous2022towards, guo2023effectively, zhang2023automatically, kimadaptive, huang2024crashtranslator, wang2024mrca, borgarelli2024reward, emam2015test, ferretti2016automatic, emam2018inferring, fu2020seads, ghaemmaghami2022automatically, gu2022correctness, moghadam2021autonomous, cardozo2023auto, ding2023integration, sherin2023qexplore, giamattei2025reinforcement, romdhana2022deep, guo2025effectively}             & \textcolor{awesome}{\textbf{27}}     \\
    &        &     & Deep Q-Network (DQN)                                    & \cite{ chen2018drlgencert, tufano2022using, zhang2023resplay, yang2023deep, zhang2023android, lan2024deeply, doreste2024adversarial, puaduraru2024end, ahmadi2022dqn, lakhan2022efficient, shu2023boosting, lu2022learning, domini2024scarlib, yu2024effective, liu2023generation, eberhart2021dialogue, giamattei2025reinforcement, joshi2024comparative, nouwou2023comparison, mindom2025harnessing, liang2025garl, bagherzadeh2021reinforcement}   & 22    \\
    &        &                     & Double Deep Q-Network (DDQN)                            & \cite{koroglu2021functional, akazaki2018falsification}             & 2     \\
    &        &                     & State-Action-Reward-State-Action (SARSA)                 & \cite{gu2017aimdroid, li2024class, almulla2022learning}              & 3     \\
    \multirow{13}{*}{\rotatebox[origin=c]{90}{\textcolor{awesome}{\textbf{MDP-based RL}}}}     &   &                     & Many-Objective Q-Learning                               & \cite{haq2023many}              & 1     \\
        &    &                     & Multi-Agent Q-Learning                                  & \cite{fan2024can, kim2024multi}              & 2     \\
        &    &                     & Deep Recurrent Q-Network (DRQN)                         & \cite{su2022effectively}             & 1     \\ 
        &    &                     & Value Iteration                        &  \cite{reddy2020quickly, kang2024macs}            & 2     \\ \cline{3-6} 
        &    & Policy-based        & Curiosity-ES & \cite{le2024exploration}              & 1     \\
         &    &        & REINFORCE & \cite{tan2024energy}              & 1     \\
        &     &                     & Vanilla Policy Gradient Technique  & \cite{ren2024prorlearn,spieker2017reinforcement, wu2019reinam, chen2019relational, liu2022learning, wang2022learning, shi2023machine, guo2024smart, yao2024carl, bertolino2020learning}              & 10     \\
        &    &                     & Proximal Policy Optimization (PPO)  & \cite{li2024quarl, romdhana2022ifrit, ding2023tracediag, corradini2024deeprest, eom2024fuzzing, gao2024preference, zhang2023challenging, de2023exploiting, humeniuk2024reinforcement, takerngsaksiri2025pytester, abadeh2024knowledge, turker2024accelerating, joshi2024comparative, nouwou2023comparison, mindom2025harnessing, qian2025reinforcement, bagherzadeh2021reinforcement}              & \textcolor{awesome}{\textbf{17}}     \\
        &    &                     & Trust Region Policy Optimization (TPRO)  &      \cite{qian2025reinforcement, bagherzadeh2021reinforcement}        & 2     \\        
        &    &                     & Self-Critical Sequence Training (SCST)                  & \cite{liu2019automatic}              & 1     \\ 
            \cline{3-6} 
            &            & Actor-Critic & Vanilla Actor-Critic & \cite{wan2018improving, li2024ircoco, huang2020towards, wang2020reinforcement}  & 
            4            \\
            &            &  & Soft Actor-Critic (SAC) & \cite{romdhana2022deep, song2023mathtt, bagherzadeh2021reinforcement}  &  3
                        \\                        
    &     & & Advantage Actor-Critic (A2C)                      & \cite{zheng2019wuji, chen2020enhanced, yu2023loop, liu2024compiler, chakraborty2024rlocator, tu2024isolating, joshi2024comparative, nouwou2023comparison, ding2025optimizing, qian2025reinforcement, bagherzadeh2021reinforcement}              & \textcolor{awesome}{\textbf{11}}     \\
     &       &                     & Multi-Agent Actor-Critic (MAAC)                & \cite{chen2023compiler}              & 1     \\  
     &       &                     & Asynchronous Advanced Actor Critic (A3C)                  & \cite{akazaki2018falsification}              & 1     \\
    &       &                     & Deep Deterministic Policy Gradient (DDPG) & \cite{zhang2021figcps, romdhana2022deep, song2023mathtt, bagherzadeh2021reinforcement}              & 4     \\ 
    &       &                     & Twin Delayed DDPG (TD3) & \cite{romdhana2022deep, song2023mathtt, bagherzadeh2021reinforcement}              & 3     \\ 
     &       &                     & Multi-Agent Deep Deterministic Policy Gradient (MADDPG) & \cite{wang2024faasconf, kang2024macs}              & 2     \\ 
     &       &                     & Actor-Critic using Kronecker-Factored Trust Region (ACKTR) &     \cite{qian2025reinforcement, bagherzadeh2021reinforcement}         & 2     \\
     &       &                     & Actor-Critic with Experience Replay (ACER) &     \cite{qian2025reinforcement, bagherzadeh2021reinforcement}         & 2     \\ 
    &        &                     & Reinforcement Learning from Human Feedback (RLHF) & \cite{nashaat2024towards}              & 1     \\ 
            \cline{3-6} 
& \textbf{Model-based} &                    &                                                        & \cite{nguyen2021rltcp}             & 1     \\ \midrule
\multirow3{*}{\rotatebox[origin=c]{90}{\textcolor{awesome}{\textbf{Bandit}}}} &   & Multi-Armed Bandit    & e.g., $\varepsilon$--Greedy, Upper Confidence Bound (UCB), and Thompson Sampling                                                    &      \cite{degott2019learning, hanna2025reinforcement, almulla2022learning, canino2018stochastic, ran2023badge, yu2024multiple}       & \textcolor{awesome}{\textbf{6}} 
\\ 
&       &        Contextual Bandit             & e.g., contextual $\varepsilon$--Greedy &       \cite{spieker2020adaptive}        & 1     
\\  \midrule
\multirow4{*}{\rotatebox[origin=c]{90}{\textcolor{awesome}{\textbf{Others}}}} &   & Meta-Reinforcement Learning    &                                                   &   \cite{sharma2022actor}        & 1 
\\ 
    &       &        The papers reference the use of RL but do not specify the particular algorithms employed.        &  &       \cite{molloy2025mecha, yang2020systematic, krupitzer2022proactive, yang2023sparse, wang2024rlcoder}        & 5     
\\ \midrule
\bottomrule
\end{tabular}
\end{adjustbox}
\end{table}

\section{RQ3: RL Algorithms Applied to SE}
In this RQ, we conduct a structured analysis of the RL algorithms adopted in SE studies by examining them across three primary classical dimensions: problem formulation, learning paradigm, and optimization strategy.
First, based on the Markov Decision Process (MDP) framework, we distinguish between \textbf{MDP-based RL} and \textbf{Bandit} algorithms, the latter assuming a static environment without state transitions.
Within MDP-based methods, the learning paradigm differentiates techniques based on whether an explicit model of the environment is constructed.
Accordingly, we categorize these into: \textbf{Model-Free RL}, where the agent learns policies directly through interaction without modeling the environment's dynamics; and \textbf{Model-Based RL}, where the agent builds or uses a model to simulate environment transitions and plan actions accordingly.
The optimization strategy defines how the RL agent learns and improves its decision-making policy. 
Based on this, we further categorize Model-free RL algorithms into three types: \textbf{Value-Based}, where the agent learns a value function to guide action selection; \textbf{Policy-Based}, which directly optimizes the policy without estimating value functions; and \textbf{Actor-Critic}, a hybrid approach that combines value estimation with policy optimization.
The remaining RL algorithms that do not fall under the aforementioned categories are grouped into an \textbf{Others} category, which includes approaches such as \textit{Meta-Reinforcement Learning}.
To systematically capture the diverse usage of RL algorithms, our classification also includes empirical studies that evaluate multiple approaches, as well as technical studies that integrate or ensemble various approaches. 
As a result, some papers are counted under multiple RL algorithm categories, reflecting the breadth of experimentation and comparison.

\textbf{Distribution of Applied RL Algorithms.}
Table~\ref{tab:tech} summarizes the reinforcement learning algorithms used in the surveyed papers.
In some instances, papers mention the use of policy gradient methods without specifying the exact algorithm; these are labeled as ``Vanilla Policy Gradient Technique''.
In other cases, where the use of RL is mentioned without any further technical details, we categorize them under ``Others''.
We first observe that SE researchers have applied a diverse set of RL algorithms to address various tasks, with the majority adopting MDP-based approaches.
In particular, 25 model-free MDP-based algorithms have been employed across the surveyed studies, underscoring the flexibility and effectiveness of this class in SE contexts.
Among the three optimization strategies under model-free methods, value-based RL is the most widely adopted, appearing in 60 papers (52\%), followed by actor-critic algorithms (34 papers) and policy-based methods (32 papers).
Within value-based methods, Q-learning stands out as the most frequently applied algorithm, used in 27 papers (45\% of value-based studies). 
The second most popular is Deep Q-Network (DQN), identified in 22 papers.
For policy-based methods, Proximal Policy Optimization (PPO) is the most commonly used, with 17 papers categorized.
Among actor-critic methods, Advantage Actor-Critic (A2C) is the most prominent, appearing in 11 papers.
We also observe that SE researchers tend to adopt more advanced RL algorithms, not limited to classical algorithms, including multi-agent RL and meta-RL.
Notably, with the rapid advancement of large language models (LLMs), integrating RL with LLMs to tackle complex SE challenges has become popular. 
For example, Reinforcement Learning from Human Feedback (RLHF) has been leveraged to align LLM outputs with developer expectations, enabling more effective handling of tasks (i.e., automated code review~\cite{nashaat2024towards}).

On the other hand, the table shows that model-based RL algorithms are rarely applied in addressing SE tasks, with only one paper~\cite{nguyen2021rltcp} identified in our survey.
Model-based RL refers to approaches in which the agent explicitly learns or is provided with a model of the environment's dynamics, that is, how actions lead to state transitions and rewards, and then uses this model to simulate future interactions and plan optimal actions. 
Although model-based methods are generally known for their high sample efficiency and strong planning capabilities in traditional RL applications, their limited presence in SE research may stem from the complexity of accurately modeling SE environments, such as code changes or software behaviors, which are often highly dynamic, domain-specific, and difficult to simulate accurately.

\begin{figure}[t]
    \centering
    \subfloat[Evolution of applied RL algorithms over time\label{fig:evolution}]{
    \includegraphics[width=.8\textwidth]{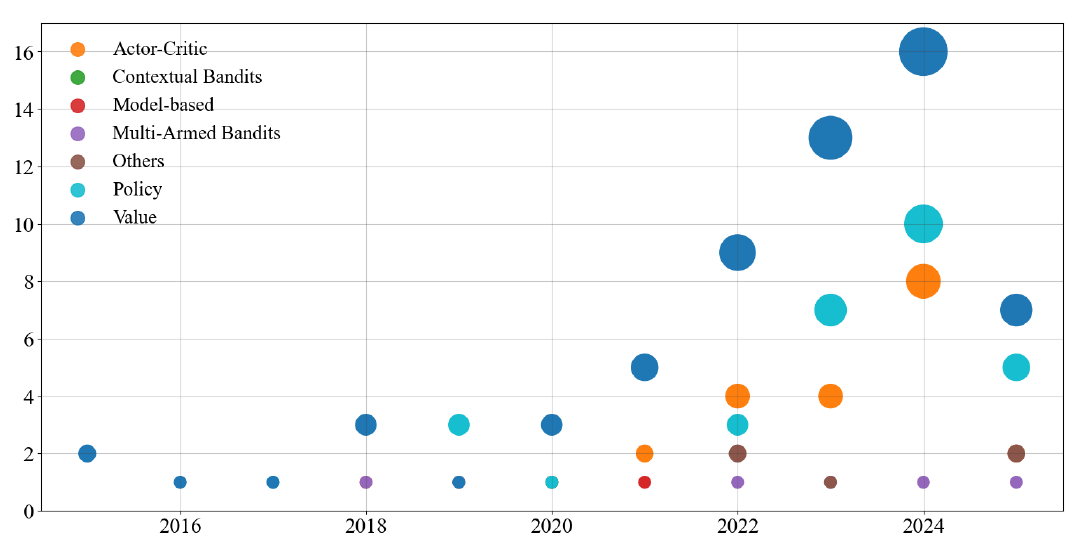}
    }\\
    \subfloat[Applied RL algorithms across SE tasks\label{fig:rlintask}]{
        \includegraphics[width=.8\textwidth]{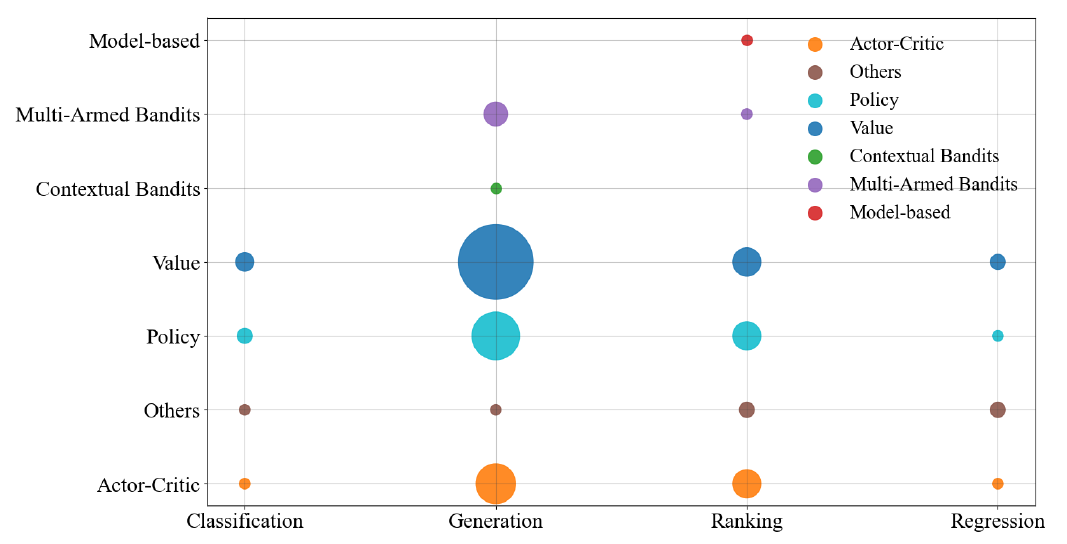}
    }
    \caption{Fine-grained analysis of RL application over time and across SE tasks}
    \label{fig:bullbleplot}
\end{figure}

Results suggest that value-based methods, particularly Q-learning and DQN, remain popular in RL-for-SE studies due to their relative simplicity, ease of implementation, and effectiveness in many SE scenarios. 
Their widespread adoption indicates that they offer a practical entry point for researchers applying RL in SE tasks. 
At the same time, the observed use of more advanced optimization techniques, such as policy-based methods and actor-critic approaches, reflects a maturing research landscape. 
This trend highlights a growing interest in leveraging more sophisticated and adaptive RL strategies to address the increasing complexity and scalability demands of real-world SE problems.
Meanwhile, the mapping results highlight the need for deeper exploration of more recent advances (e.g., SAC, TD3, meta-RL, multi-agent RL, RLHF).
These methods offer enhanced stability, sample efficiency, and adaptability, which could be particularly valuable for addressing complex and dynamic SE tasks that go beyond what traditional RL algorithms can effectively handle.
Furthermore, our manual classification reveals that a portion of the surveyed papers did not explicitly specify the RL algorithms used. 
This lack of clarity can hinder reproducibility and impede rigorous comparison across studies. 
Therefore, we strongly encourage future studies to clearly report the specific RL algorithms, which would facilitate more transparent benchmarking, better understanding of design decisions, and easier replication of results.

\textbf{Time-Based Analysis of RL Algorithm Adoption in SE.}
We further present a bubble plot in Figure~\ref{fig:evolution} to visually illustrate the distribution and evolution of applied RL algorithms across the surveyed time frame. 
As shown in the figure, value-based RL algorithms have been consistently adopted since 2015 and have increasingly emerged as the dominant approach for addressing SE tasks. Notably, their usage peaked in 2024, with 16 surveyed papers. 
This trend reinforces our earlier findings that value-based methods are favored for their relative simplicity, ease of implementation, and proven effectiveness across a variety of SE scenarios.
Policy-based methods first appeared in the SE domain in 2019, while actor-critic methods were introduced later in 2021. 
Similarly, the number of surveyed papers employing these two algorithms peaked in 2024, with 10 papers utilizing policy-based methods and 8 papers adopting actor-critic methods.
It is observed that both algorithms have gradually gained popularity and have been increasingly applied in SE scenarios since 2021, reflecting the growing interest in leveraging more sophisticated RL strategies for complex software engineering tasks.

\textbf{Relationship Between RL Algorithms and SE Task Types.}
The bubble plot shown in Figure~\ref{fig:rlintask} illustrates the distribution of various RL algorithms across the four SE tasks (i.e., classification, generation, ranking, and regression) as categorized in R2.
First of all, we observe that Bandit-based methods are exclusively applied in generation and ranking tasks. 
This may be attributed to their suitability for scenarios involving immediate reward feedback and limited environmental dynamics, which align well with tasks such as output generation and prioritization.
The figure also reveals that different SE tasks tend to favor specific types of RL algorithms, suggesting a natural alignment between the characteristics of each task and the strengths of particular RL paradigms.
For example, in the generation task, value-based methods are the most widely applied, followed by policy-based and actor-critic approaches.
Generation tasks often involve discrete action spaces (e.g., token or code snippet selection), where value estimation can efficiently guide the construction of outputs. Policy-based and actor-critic methods, while more complex, are increasingly adopted due to their ability to handle continuous outputs and provide better stability in long-horizon generative processes.
For the remaining three SE tasks (i.e., classification, ranking, and regression), no single RL algorithm type dominates. 
This may indicate that the choice of RL algorithms in these tasks remains exploratory, possibly due to the smaller number of studies and the diverse nature of problem settings that do not yet favor one particular approach.

\begin{tcolorbox}[colback=gray!5,colframe=awesome,title= RQ3 Summary]
\textcolor{BurntOrange}{\faStar}~Through a comprehensive analysis of 115 RL-for-SE studies, we identify a strong preference for model-free, value-based methods, highlighting their practicality and effectiveness in SE contexts. 
At the same time, an emerging trend toward advanced RL strategies (e.g., policy-based, actor-critic, multi-agent RL, and RLHF) reflects the community's response to increasingly complex SE challenges.\\
\textcolor{BurntOrange}{\faStar}~Our temporal analysis shows that value-based methods have been consistently utilized since 2015, while policy-based and actor-critic algorithms have gained traction since 2021.\\
\textcolor{BurntOrange}{\faStar}~In terms of SE task types, generation tasks exhibit a strong reliance on value-based approaches, whereas classification, ranking, and regression tasks show more varied algorithm usage and remain comparatively underexplored.
\end{tcolorbox}

\section{RQ4: Design and Evaluation Practices in RL-for-SE tasks}
In this RQ, we performed an in-depth analysis to investigate how the surveyed RL-for-SE papers design the tasks...

\textbf{Dataset Sources.}
\begin{figure}
    \centering
    \includegraphics[width=.7\linewidth]{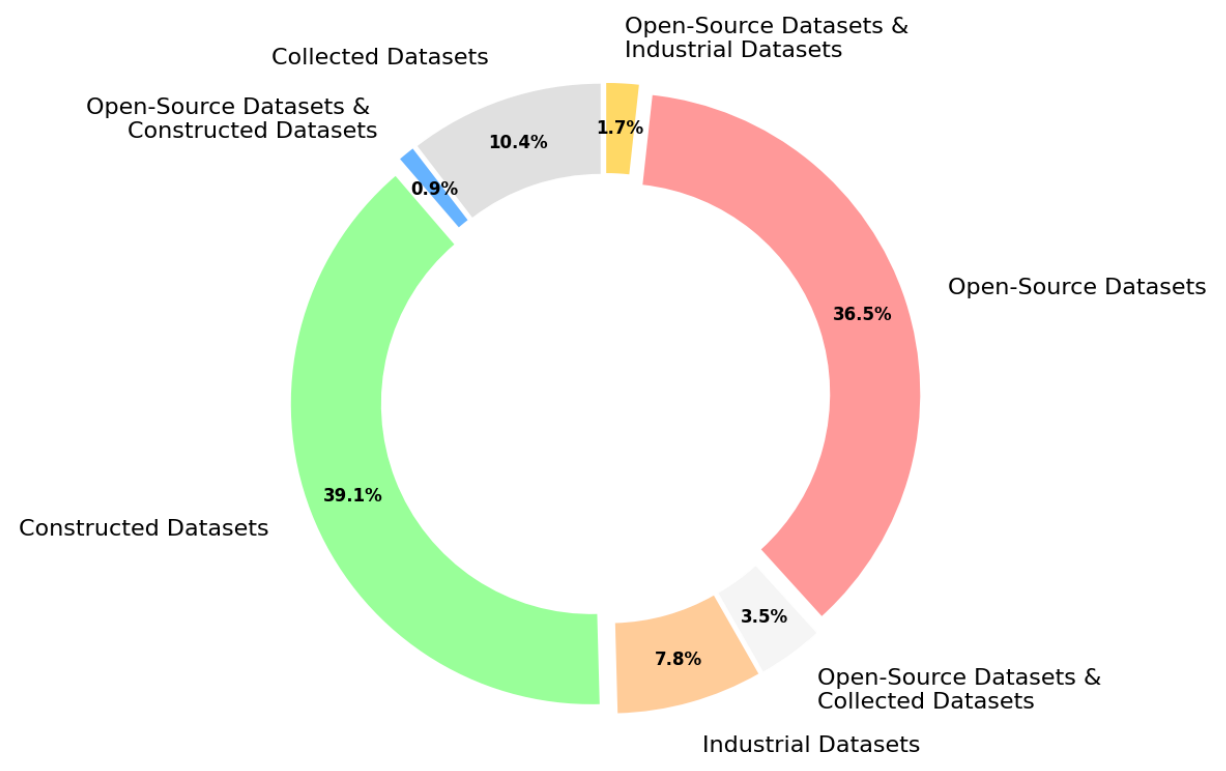}
    \caption{Distribution of data sources in RL-for-SE tasks}
    \label{fig:enter-label}
\end{figure}
Data plays a crucial role in training models, as it directly influences their generalization ability, effectiveness, and overall performance. 
Therefore, in the context of data, we examine how the surveyed papers collect their datasets and identify the sources from which the data is obtained.
Following the prior work~\cite{yang2022survey, hou2024large}, we specifically divide the data sources into four primary categories: Open-source Datasets, Collected Datasets, Constructed Datasets, and Industrial Datasets.  
Open-source Datasets refer to publicly accessible collections of data that are typically shared through open platforms or repositories, including well-established benchmarks, such as Defects4J and SWE-bench. 
Collected Datasets are those compiled directly by researchers from various sources, including but not limited to major websites, forums, software systems, and social media platforms, such as GitHub.
Constructed Datasets are customized datasets developed by researchers either by modifying or augmenting existing collected data, or by generating simulated data (e.g., for cyber-physical systems or autonomous driving systems), to better align with specific research objectives.
Industrial Datasets refer to data obtained from commercial or industrial sources, often containing proprietary business information, user behavior logs, and other sensitive content. These datasets are particularly valuable for research targeting real-world business scenarios.

Figure~\ref{fig:enter-label} presents the distribution of data sources used across various RL-for-SE tasks. 
It clearly illustrates that Constructed Datasets and Open-Source Datasets are the two most commonly utilized types, accounting for 39.1\% and 36.5\% of the total, respectively. This trend highlights researchers' strong preference for either tailoring datasets to specific problem domains or leveraging established benchmarks and publicly available datasets to ensure comparability and reproducibility.
Interestingly, the results reveal that only 10.4\% of the surveyed papers utilized collected datasets to train their models, substantially lower than the 33\% reported in the context of DL-for-SE tasks~\cite{yang2022survey}. 
This notable gap suggests that, in RL-for-SE research, there may be greater reliance on curated or benchmark datasets, potentially due to the higher complexity and overhead associated with preparing high-quality task-specific data suitable for reinforcement learning particularly in areas such as reward design and environment simulation.
Furthermore, real-world industrial datasets are used in only 7.8\% of the surveyed studies, indicating a potential misalignment between the datasets commonly employed in academic research and those encountered in real-world industrial contexts. 
This divergence underscores the need for future research to explore and incorporate industrial datasets. 
Doing so would help ensure that RL algorithms are not only effective in controlled academic settings but also applicable, reliable, and robust in practical, large-scale industrial scenarios.
Meanwhile, we observe that a few surveyed papers utilize multiple datasets for training and evaluation. Specifically, 3.5\% of the studies use a combination of Open-Source Datasets and Collected Datasets, 1.7\% combine Open-Source and Industrial Datasets, and 0.9\% use both Open-Source and Constructed Datasets. 
This indicates a limited but noteworthy effort to diversify data sources, which could contribute to more comprehensive and generalizable evaluations.

\begin{figure}
    \centering
    \includegraphics[width=0.6\linewidth]{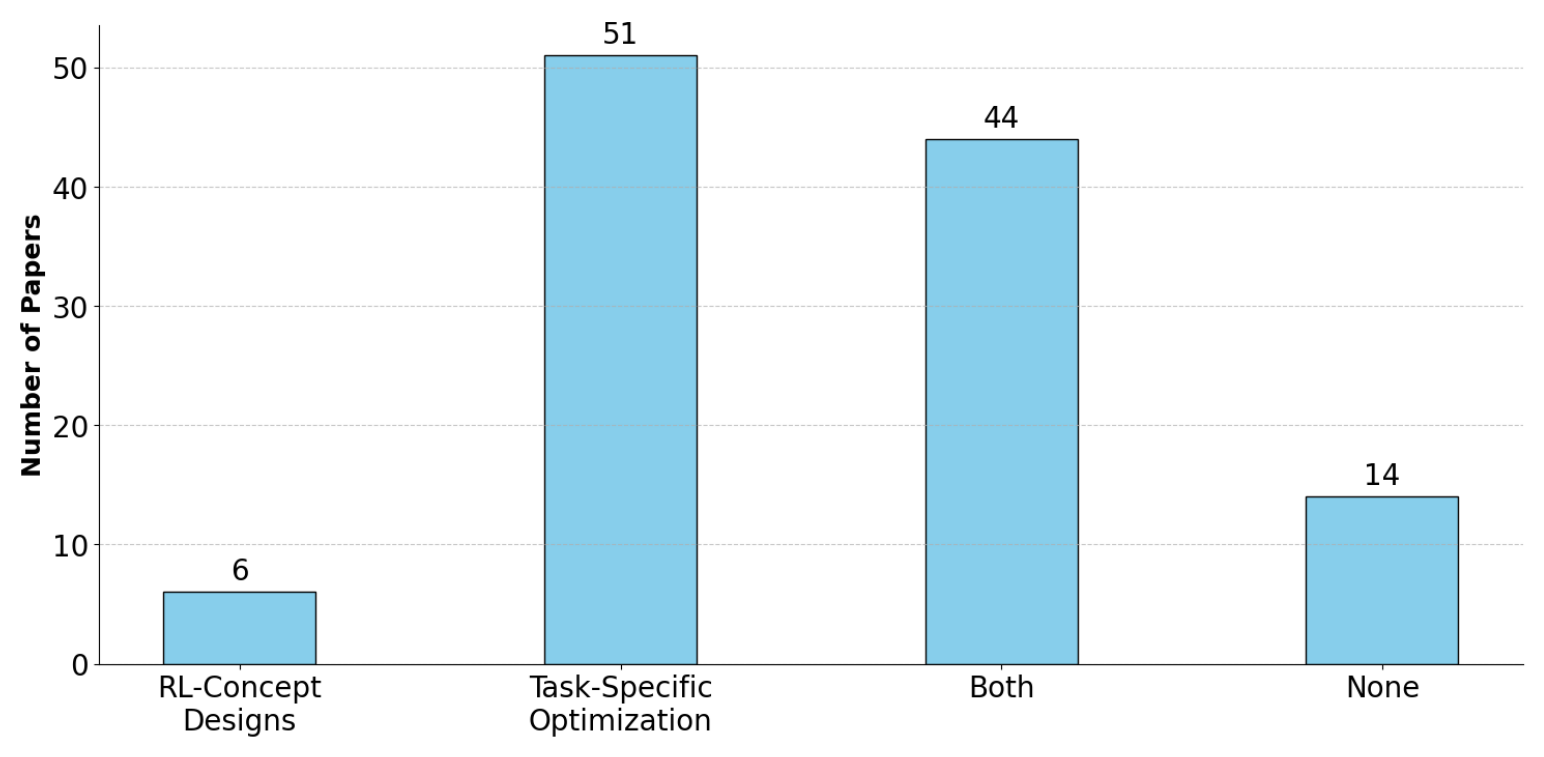}
    \caption{Distribution of used strategies in surveyed papers}
    \label{fig:methods}
\end{figure}

\textbf{Model Design and Optimization Techniques.} 
To effectively tailor RL to SE tasks, researchers often refine model architectures and apply a range of optimization techniques aimed at enhancing performance, improving convergence stability, and addressing task-specific constraints. 
To capture current practices in this area, we manually reviewed the surveyed papers and systematically annotated the optimization strategies employed across different SE applications by utilizing the card-sorting method. 
Recognizing the distinctive demands of RL-based solutions, we categorized these strategies into two overarching groups: (1) \textbf{RL-Concept Designs}, which focus on modifications to the core RL algorithms or training processes (i.e., state, action, and reward), and (2) \textbf{Task-Specific Optimization Techniques}, which incorporate domain knowledge, heuristics, or application-driven constraints to adapt the learning process to particular SE problems.
Figure~\ref{fig:methods} presents the distribution of the RL-Concept Designs and Task-Specific Optimization Techniques adopted across the surveyed studies.
To provide a more granular view, Table~\ref{tab:approaches} summarizes the specific strategies employed within these categories, accompanied by representative examples and the number of papers that utilize them.

\begin{table}[t]
\caption{Model design and optimization techniques in RL-for-SE papers}
\label{tab:approaches}
\begin{adjustbox}{max width=1.0\textwidth, center}
\begin{tabular}{cllp{5cm}r}
\toprule
\multicolumn{1}{l}{Category} & \multicolumn{2}{l}{Subcategory} & Examples & Total \\ 
\midrule

\multirow{8}{*}{\rotatebox[origin=c]{90}{\textcolor{awesome}{\textbf{RL-Concept-Designs}}}} & \multirow{3}{*}{\textbf{State}} & State Merge & State Similarity Analysis & 4 \\
 &  & State Abstraction & DNN-based State Presentation & 18 \\
 \cline{2-5}
 & \multirow{2}{*}{\textbf{Action}} & Action Supplementation & Using No-operation & 1 \\
 &  & Action Space Management & Hierarchical Action Space & 3 \\
  \cline{2-5}
 & \multirow{3}{*}{\textbf{Reward}} & Punishment Shaping & Penalty & 8 \\
 &  & Multi-level Reward & \makecell[l]{API-level+Activity-level+Widget-level\\ Reward} & 10 \\
 &  & Task Specific Reward & Immediate Reward, Cumulative Reward, Coverage-Weighted Reward & 15 \\
 \midrule
\multirow{15}{*}{\rotatebox[origin=c]{90}{\textcolor{awesome}{\textbf{Task-Specific-Techniques}}}} & \multirow{7}{*}{\textbf{Exploration}} & Random-based Exploration & \makecell[l]{$\epsilon$-greedy, Random Exploration,\\ Pseudo-random Proportional Rule} & 46 \\
 &  & Uncertainty-based Exploration & Thompson Sampling, Upper Confidence Bound & 9 \\
 &  & Curiosity-driven Exploration & Random Network Distillation, Intrinsic Curiosity Module & 14 \\
 &  & Heuristic-based Exploration & \makecell[l]{Value-difference-based Exploration, \\ Self-exploration Training} & 8 \\
 &  & Probability-based Exploration & \makecell[l]{Softmax, Temperature Softmax, \\ Gumbel-Softmax} & 6 \\
 &  & Diversity-guided Exploration & \makecell[l]{Dynamic Exploration State Reset, \\ Archive, Online Cover Strategy} & 8 \\
 &  & Safety-guided Exploration & Safe Exploration Strategy & 1 \\
 \cline{2-5}
 & \multirow{6}{*}{\textbf{Optimization-Training}} & Learning Efficiency Optimization & \makecell[l]{AdaGrad, Distributed Data Parallel, \\Minibatch Update, Momentum Update} & 19 \\
 &  & Performance Enhancement Optimization & Self-Optimizer, Max Update Rule & 27 \\
 &  & Overfitting Prevention Optimization & \makecell[l]{Early Stop, k-fold Cross-Validation \\ KL Regularization} & 9 \\
 &  & Adaptation Optimization  & \makecell[l]{Continual Learning, Fine-tuning \\ Transfer Learning, Prompt Tuning} & 13 \\
 &  & Generalization Optimization & \makecell[l]{Adversarial Training, Huber Loss, \\Counterexample-Guided Learning} & 6 \\
 &  & Feedback-Driven Optimization & \makecell[l]{Human-in-the-Loop Workflow, \\ Iterative Feedback Learning} & 4 \\
 & \multicolumn{1}{l}{\textbf{Optimization-Dataset}} & Dataset Optimization & \makecell[l]{Random Over Sampling, Random \\ Under Sampling, Data Filtering} & 7 \\
 & \multicolumn{1}{l}{\textbf{Optimization-Other}} & Other & \makecell[l]{Action Label Learning, Ensemble,\\Multiple Q-tables Management} & 16 \\
 \bottomrule
\end{tabular}
\end{adjustbox}
\end{table}

As shown in Figure~\ref{fig:methods}, among the 115 reviewed studies, only six papers (5.2\%) concentrate solely on refining the core components of RL, focusing on \textit{conceptual innovations in RL design}. In contrast, 51 papers are dedicated exclusively to applying RL to specific SE tasks, using \textit{task-specific optimization strategies}. Additionally, 44 studies adopt a hybrid approach, incorporating both RL concept refinement and task-oriented optimization. These results indicate that the vast majority of current research (82.6\%) utilizes RL primarily as a tool to address practical challenges in SE, rather than to advance the theoretical foundations of RL itself. Notably, 14 studies (12.2\%) do not clearly describe either the design of their RL methods or the optimization strategies employed, suggesting a need for improved methodological standardization in the field.

Table~\ref{tab:approaches} presents a detailed taxonomy of the reviewed studies based on their methodological focus.
The \textbf{RL-Concept Designs} category includes studies that aim to refine the core components of RL algorithms, particularly in terms of \textit{State}, \textit{Action}, and \textit{Reward} representations.

\begin{itemize}
\item \textbf{State} representation is a critical focus area, especially in practical SE tasks, which often involve complex and heterogeneous elements. For example, GUIs typically consist of various widgets and buttons, making accurate and efficient state modeling particularly challenging. To address this, \uline{state abstraction} has been proposed. Pan et al.~\cite{pan2020reinforcement} proposed using DNN (e.g., Graph Neural Networks~\cite{li2024quarl} and Bi-LSTMs~\cite{zhang2023challenging}) to abstract high-dimensional UI states into low-dimensional tuples, facilitating more manageable learning. Another challenge is that different elements within a state may exhibit varying configurations. For example, GUI widgets often possess different properties, and treating each variation as a separate state can result in a significant state space explosion during exploration. To address this, \uline{state merging} methods have been proposed. Koroglu et al.~\cite{koroglu2018qbe} introduced a cosine similarity-based approach to assess state equivalence, enabling the merging of similar states. Similarly, Zheng et al.~\cite{zheng2021automatic} proposed merging web pages that share the same business logic but differ in content details into the same state, thereby reducing the overall complexity of the state space.

\item \textbf{Action} design is also critical for SE tasks. Since SE data are often collected from real-world applications, they may be incomplete or unstructured. For example, bug reports frequently suffer from missing information due to developer oversight or limited technical background. These issues should be considered during action design. To address such cases, \uline{supplying specific actions} has been proposed. Zhang et al.~\cite{zhang2023automatically} proposed a \textit{no-operation} (no-op) action to bridge gaps caused by missing steps when aligning bug reproduction steps with UI events.
Another common challenge involves the large size of the action space, which significantly increases the difficulty of effective policy learning. Therefore, we need to \uline{manage the action space}. For example, in the task of quantum circuit optimization, a circuit with thousands of gates can lead to a combinatorial explosion, potentially resulting in millions of possible actions. To reduce this complexity, Li et al.~\cite{li2024quarl} decomposed the action space into two subspaces, i.e., position space and transformation space, which effectively lowers the decision-making complexity and reduces the overall search space.

\item \textbf{Reward} plays a critical role in guiding RL agents toward desirable behaviors in SE tasks. Some works focus on \uline{punishment shaping}, assigning penalties to discourage undesirable actions. For example, to prevent the agent from exploring invalid execution paths, Koo et al.~\cite{koo2019pyse} proposed assigning a high penalty during symbolic execution when an action leads to an infeasible path condition (i.e., no concrete input values can satisfy the path constraint). This strategy helps the agent avoid such invalid actions that prematurely terminate execution, thereby encouraging the learning of branching policies that tend to generate longer (worst-case) execution paths.
Other studies address tasks involving multiple objectives, where a \uline{multi-level reward} is designed to balance competing goals. For example, for game testing, Zheng et al.~\cite{zheng2019wuji} defined a reward that encourages agents to explore diverse game scenarios and to win the game, considering both exploration and performance. Similarly, Zhang et al.~\cite{zhang2023android} tackled GUI testing by designing three levels of rewards, i.e., API-level, activity-level, and widget-level, aimed at encouraging the agent to explore more activities that contain sensitive APIs as well as discover new UI states.
In some cases, \uline{task-specific rewards} are designed for domain characteristics. For example, in test case prioritization for regression testing, Yang et al.~\cite{yang2020systematic} proposed a history-based reward, under the assumption that test cases that previously exposed faults are more likely to expose faults again. This reward design guides the agent to prioritize historically effective test cases during learning.

\end{itemize}

In contrast, the \textbf{Task-Specific Techniques} category includes a wide range of strategies designed for SE tasks. This category can be further divided into two subcategories: exploration and optimization. 
For each subcategory, we highlight and discuss the most frequently adopted techniques.
\begin{itemize}
\item \textbf{Exploration} includes seven distinct types of methods, among which the most widely used are random-based (mentioned in 46 papers), curiosity-driven (14 papers), and uncertainty-based approaches (9 papers).
\uline{Random-based exploration} introduces randomness to action selection, forcing the agent to explore unknown areas. For example, $\epsilon$-Greedy balances exploration and exploitation by selecting random actions with a probability of epsilon. Decayed $\epsilon$-Greedy gradually reduces epsilon as training progresses, favoring exploration in the early stages and exploitation in the later stages.
\uline{Curiosity-driven approaches} encourage agents to explore novel states by simulating \textit{curiosity} about unknown states. For example, Zhang et al.~\cite{zhang2021figcps} used Random Network Distillation to model curiosity through prediction errors of a fixed random target, guiding their approach to explore rare unsafe states in cyber-physical software.
\uline{Uncertainty-based approaches} leverage uncertainty (or variance) in value estimation to guide exploration. For example, in UI testing tasks, it is essential to automatically learn the acceptable interactions for different UI elements (e.g., buttons and text boxes). Degott et al.~\cite{degott2019learning} applied Thompson Sampling to dynamically handle uncertainty in UI interactions, enabling more accurate identification of effective interactions and reducing blind exploration.

\item \textbf{Optimization-Training} includes a wide range of approaches aimed at optimizing the training process. The most widely used strategies focus on improving performance (27 papers), enhancing learning efficiency (19 papers), and adapting to new tasks (13 papers). \uline{Performance enhancement} is particularly critical for SE tasks. For example, in testing tasks for complex systems such as distributed systems, which often involve sparse rewards, Borgarelli et al.~\cite{borgarelli2024reward} applied the Max Update Rule to allow RL algorithms to explore the state space more effectively. This approach maximizes the optimal balance between the exploration rewards for new states and potential future rewards. By prioritizing the coverage of unvisited states, it significantly improves test coverage and bug-finding capability. Yang et al.~\cite{yang2023deep} used multi-round training, which iteratively cycles through ``single-round training success $\to$ experience collection $\to$ function updates $\to$ next-round training optimization.'' This process enables the model to accumulate effective experience across rounds, thereby improving the efficiency of program synthesis.
\uline{Improving learning efficiency} is equally important. For example, in code summarization tasks, training involves the fusion of multiple code features, which leads to significant differences in parameter update frequency and gradient magnitude. To address this, Wang et al.~\cite{wang2020reinforcement} used Adagrad to dynamically adjust the learning rate, improving training stability and efficiency. Similarly, Li et al.~\cite{li2024quarl} leveraged Distributed Data Parallel to accelerate the pretraining process. By fully utilizing multi-GPU resources, this approach enhances model training efficiency.
Moreover, \uline{smooth adaptation} to SE tasks is crucial. For example, Doreste et al.~\cite{doreste2024adversarial} employed continual learning to enable ego ADS (autonomous driving systems under test) to retain existing driving capabilities while learning new strategies to counter adversarial ADS. P{\u{a}}duraru et al.~\cite{puaduraru2024end} used pretrained models to initialize DQN, integrating human behavioral knowledge. This not only accelerates adaptation to new tasks but also allows the model to surpass human performance.

\item \textbf{Optimization-Dataset} focuses on improving the quality of datasets to ensure effective training. For example, when dealing with the imbalance in pull-request analysis datasets, Joshi et al.~\cite{joshi2024comparative} employed random under-sampling and over-sampling techniques to achieve dataset balance. In test case prioritization tasks, Ding et al.~\cite{ding2023integration} utilized the stubbing complexity and execution complexity of each test case to guide prioritization. Since these metrics had inconsistent dimensions, normalization was applied to enable accurate calculations of reward.
Additionally, Nashaat et al.~\cite{nashaat2024towards} proposed augmenting datasets through an iterative generation and filtering mechanism based on self-instruct. This method generates new and diverse training samples from the initial data, thereby enriching the instruction dataset used for fine-tuning language models. 

\end{itemize}

\textbf{Employed Evaluation Metrics.} 
\begin{table}[]
\caption{Proportion of evaluation metrics employed in surveyed papers}
\label{tab:metric}
\begin{threeparttable}
\begin{tabular}{lp{8cm}c}
\toprule
\textbf{Metric Category} & \textbf{Examples} & \textbf{Percent} \\ \midrule
\textbf{Effectiveness}     &   Accuracy, Precision, Recall, F1, Coverage, AUC, ROC, Unique Faults, Success Rate, Failure Rate, Top-N, Quality, BLEU, BLEUS-4, METEOR, ROUGE-L, CodeBLEU, Exact Match, Edit Similarity, Mean Reciprocal/First/Average Rank, Mutation Score     & \blackwhitebar{0.49}
  \\
\textbf{Efficiency} &     Running Time, Training time, Inference Time, Computational Cost, Gas Consumption, Memory           & \blackwhitebar{0.06}  \\
\textbf{Effectiveness \& Efficiency} &              & \blackwhitebar{0.44}  \\
\bottomrule
\end{tabular}
\end{threeparttable}
\end{table}
In evaluating RL-for-SE studies, we manually reviewed and annotated the evaluation metrics employed in each work. 
To enable a structured analysis, we grouped these metrics into two key categories: (1) Effectiveness, which captures how well the RL-based approach performs in achieving its intended task (e.g., accuracy, precision, success rate, and BLEU), and (2) Efficiency, which reflects the computational or resource-related aspects of the approach (e.g., running time, memory, and training time).
By considering both dimensions, we aim to assess not only the performance of RL techniques but also their practicality and scalability in real-world software engineering scenarios.
In particular, we are interested in the extent to which existing studies address efficiency, as this remains a critical yet often underreported factor in the adoption of RL in SE tasks.
Table~\ref{tab:metric} shows the proportion of evaluation metrics employed in surveyed papers.

As shown in the table, fewer than half of the surveyed papers (44\%) consider both effectiveness and efficiency in their evaluations, while 49\% focus solely on effectiveness. 
This highlights a notable gap in assessing the practical applicability of RL approaches in software engineering. Moreover, our manual analysis reveals that among those addressing efficiency, most studies primarily consider running time and inference time, with limited attention to other important aspects such as training time, memory consumption, and scalability. 
Meanwhile, the table shows that approximately 6\% of the surveyed papers focus solely on efficiency-related metrics. These studies typically pertain to software design tasks, such as load balancing of controllers in software-defined networks, where resource utilization and performance overhead are of primary concern.

\textbf{Replicability of surveyed papers.} We assess the replicability of RL-for-SE studies by examining the availability of their source code. Specifically, we first identified whether a replication package (e.g., hosted on GitHub, Zenodo, or Figshare) was provided. 
We then manually inspected these links to verify whether the actual source code was accessible. 
Table~\ref{tab:repli} shows the proportion of replicability for surveyed papers.
The analysis reveals that a substantial proportion (41\%) of the surveyed papers are not replicable, making it difficult for other researchers to reproduce their results. 
These studies either lack a replication package entirely, provide broken links (e.g., 404 errors), or offer only partial resources (such as supplementary results or datasets, without including the source code).
Compared to conference papers, journal papers exhibit lower applicability, with 46\% being non-replicable, whereas 36\% of conference papers lack replicability.

\begin{table}[]
\caption{Proportion of replicability for surveyed papers}
\label{tab:repli}
\begin{threeparttable}
\begin{tabular}{lccc}
\toprule
               & \textbf{Conference Papers} & \textbf{Journal Papers} & \textbf{Percent} \\ \midrule
\textbf{Replicable}     & \blackwhitebar{0.64}             & \blackwhitebar{0.54}          & \blackwhitebar{0.59}
  \\
\textbf{Non-Replicable} & \blackwhitebar{0.36}             & \blackwhitebar{0.46}           & \blackwhitebar{0.41}  \\ \bottomrule
\end{tabular}
\end{threeparttable}
\end{table}

Based on this observation, obtaining open-source implementations of RL-based approaches remains a significant challenge in SE. 
Unlike traditional ML or DL methods, RL-based approaches are inherently more complex due to their interactive learning loop, dynamic feedback mechanisms, and task-specific reward designs. Moreover, RL-based SE studies often involve diverse data preprocessing pipelines, specialized representation strategies, and customized optimization techniques, all of which can be difficult to document and standardize. 
Therefore, we strongly recommend that future RL-for-SE studies consistently release comprehensive replication packages, including all necessary source code, datasets, and configuration files, to ensure reproducibility and facilitate further research advancement.

\begin{tcolorbox}[colback=gray!5,colframe=awesome,title= RQ4 Summary]
\textcolor{BurntOrange}{\faStar}~Our in-depth analysis of RL-for-SE studies reveals strengths and limitations in current research practices. In terms of dataset usage, we find a strong reliance on Constructed and Open-Source Datasets, with limited adoption of Collected and Industrial Datasets, highlighting a need for broader data diversity and stronger alignment with real-world scenarios.\\
\textcolor{BurntOrange}{\faStar}~Regarding model design and strategies used, most studies prioritize task-specific optimization strategies over innovations in core RL algorithms, indicating that RL is often used as a tool to address SE challenges rather than to advance RL theory.\\
\textcolor{BurntOrange}{\faStar}~Evaluation practices also show a predominant focus on effectiveness, with efficiency considerations remaining underreported.\\
\textcolor{BurntOrange}{\faStar}~Moreover, replicability remains a significant challenge, with over 40\% of studies, especially those published in journals, lacking accessible implementation resources.
\end{tcolorbox}

\section{RQ5: Challenges and Opportunities}
\label{sec:chall}
Despite growing interest in RL-for-SE, several important research challenges and opportunities remain underexplored. 
Building on the insights and observations from our study, this section outlines key limitations in the current body of RL-for-SE research. We summarize six open challenges that require deeper investigation and present six actionable recommendations to guide future work in this evolving field.

\subsection{Challenges}
We identify and elaborate on six key challenges that currently hinder the broader adoption and advancement of RL-for-SE.
These challenges are outlined and discussed as follows:

\textcolor{red}{\faFrownO}~\textbf{Challenge 1 - Lack of Empirical Comparison and User-Centered Validation in RL-for-SE.} 
Although the RL-for-SE landscape is dominated by studies introducing new techniques or methodologies, comprising approximately 88\% of surveyed papers, there is a noticeable shortfall in empirical rigor and user-centered evaluation. 
Most studies focus on presenting a single RL-based solution without sufficiently comparing it against alternative RL algorithms or baseline methods within the same task. 
As a result, it becomes difficult to assess the relative advantages, limitations, and trade-offs of different RL algorithms when applied to specific SE scenarios.
Furthermore, the paucity of user studies and practical field deployments indicates a disconnect between academic innovation and real-world software engineering practices. Without robust empirical comparisons and human-in-the-loop evaluations, the practical feasibility, stability, and developer acceptance of these RL approaches remain unclear, posing a significant barrier to their adoption.

\textcolor{red}{\faFrownO}~\textbf{Challenge 2 - Overconcentration on SQA-Test Generation Limits Broader Impact.} Despite the promise of RL in SE, our literature review reveals a key challenge lies in the current imbalance in its application across SE activities, with the majority of RL-for-SE efforts concentrated in software quality assurance (72\% of surveyed papers), particularly test generation (49 papers being identified).
Other critical areas like software requirements, design, and management remain largely underexplored. 
This imbalance not only limits the generalizability of current findings but also restricts the broader impact RL can have across the full software lifecycle. 
One core reason for this disparity is potentially the lack of well-defined reward functions and environments in non-testing domains. Unlike testing tasks, where success metrics such as code coverage or test pass rates offer direct feedback signals, domains such as requirements engineering, architectural design, or team management involve subjective, multi-objective, or delayed outcomes, making it difficult to formulate effective RL environments.

\textcolor{red}{\faFrownO}~\textbf{Challenge 3 - Underutilization of Advanced RL Algorithms.} 
Despite growing interest in applying RL to SE tasks, the majority of existing works still rely on basic RL algorithms, such as Q-learning (23.4\%), DQN (19.1\%), PPO (14.8\%), and A2C (9.6\%). These methods were originally proposed between 1992 and 2017 and do not fully reflect the significant advancements that have emerged in the RL community in recent years. In contrast, more advanced RL algorithms, e.g., Decision Transformer~\cite{chen2021decision}, Gato~\cite{reed2022generalist}, GRPO~\cite{shao2024deepseekmath}, or SayCan~\cite{ahn2022can}, have shown strong performance in domains like robotics and language modeling by incorporating large pre-trained models, structured reasoning, and multimodal inputs. These approaches remain largely underexplored in SE applications.
Furthermore, recent algorithmic variants tailored for practical challenges could be better suited for SE tasks. For example, Implicit Q-Learning (IQL)~\cite{kostrikov2021offline} introduces stable, offline-friendly policy learning, making it well-aligned with SE tasks where only historical data (e.g., commit logs or bug-fix histories) are available. Similarly, Safe-PPO~\cite{gu2024review} extends PPO with Lagrangian-based constraint handling, which is promising for safety-critical SE applications such as automated patch generation or configuration optimization for safety-critical software with constraints.
Current RL-for-SE research is still limited by its reliance on basic RL algorithms, while many modern RL advances remain untapped, representing a key challenge.

\textcolor{red}{\faFrownO}~\textbf{Challenge 4 - The Data and Reproducibility Bottleneck.} A major challenge in advancing RL-for-SE research stems from the lack of diverse, realistic, and reusable datasets.
While many studies rely on constructed or open-source datasets, the scarcity of collected and industrial datasets restricts the ecological validity and generalizability of proposed approaches, further likely exacerbating the disconnect between academic research and practical deployment.
Constructing realistic environments for SE tasks, especially those involving reward design or dynamic interaction, remains non-trivial and often requires substantial domain knowledge and engineering effort. 
Moreover, the field suffers from low replicability: 41\% of the surveyed works are non-replicable due to missing source code, broken links, or incomplete replication packages. 
This lack of transparency not only hinders reproducibility and fair comparison but also slows the overall progress of the community. 
Alarmingly, we found that 14 studies (12.2\%) failed to clearly specify the RL algorithms used or the optimization strategies applied, further complicating efforts toward methodological standardization.
Without standardization and open-access datasets spanning diverse SE contexts, RL algorithms risk being validated primarily in constrained or synthetic settings rather than real-world scenarios.

\textcolor{red}{\faFrownO}~\textbf{Challenge 5 - Rethinking RL-for-SE Through Conceptual Innovation.}
Our analysis reveals that the majority of RL-for-SE studies predominantly emphasize adapting reinforcement learning to fit specific software engineering tasks through customized optimization strategies, while paying relatively little attention to advancing RL's core conceptual components, such as state abstraction, action space structuring, and reward formulation. 
Only 5.2\% of the papers focus solely on RL-concept refinement, whereas over 82\% either exclusively or primarily adopt task-specific techniques.  While these optimizations are effective for specific SE tasks, they often overlook the core principles and generalizable methodologies of RL. 
This reflects a trend where RL is treated more as a plug-in tool rather than a methodology to be innovatively enhanced. Such reliance on task-driven adaptations can limit the generalizability and theoretical contributions of RL-for-SE research. 
Addressing this challenge requires a stronger emphasis on principled RL algorithm design that not only meets the needs of individual tasks but also contributes reusable insights and methods to the broader RL community.

\textcolor{red}{\faFrownO}~\textbf{Challenge 6 - Limited Consideration of Efficiency in RL-for-SE Evaluation.}
Despite RL's substantial computational overhead, the majority of RL-for-SE studies primarily focus on effectiveness metrics, with limited attention given to efficiency-related concerns. 
Our analysis reveals that fewer than half of the surveyed works evaluate both dimensions, and among those that address efficiency, most restrict their scope to basic runtime or inference time measurements.
Moreover, even within well-studied SE tasks, such as test case generation, there is a noticeable lack of consistency in evaluation practices. 
Many studies adopt task-specific or ad hoc metrics, and some entirely overlook efficiency considerations. 
This lack of standardized, comprehensive evaluation not only limits cross-study comparability but also weakens the assessment of RL's practicality in real-world SE scenarios.

\subsection{Opportunities}
Although several challenges remain unresolved, the rapid advancement of technologies, as well as the insights accumulated from prior studies, present promising opportunities to advance future RL-for-SE research. 
Six major opportunities are outlined as follows:

\textcolor{teal}{\faSmileO}~\textbf{Opportunity 1 - Empirical Benchmarking and Human-Centric Design for Practical RL-for-SE.}
The current method-centric focus creates a ripe opportunity for more comparative and human-grounded studies. 
First, conducting in-depth empirical evaluations of multiple RL algorithms on the same SE task (e.g., test case generation, bug localization, or code optimization) can offer critical insights into which methods perform better under specific constraints such as data sparsity, reward sparsity, or system dynamics. This would not only help identify the most suitable RL strategies for different categories of SE tasks, but also build a foundation for reusable evaluation protocols and fair benchmarking across future work.

Second, the field would benefit from stronger user-centered and practice-driven studies. Integrating RL solutions into real development workflows, paired with developer feedback, usability studies, or observational research, could reveal practical frictions such as interpretability barriers, trust issues, or integration bottlenecks. These findings would be instrumental in crafting RL systems that are not only technically effective, but also usable, explainable, and sustainable.
Moreover, introducing user-centered evaluations, such as usability testing, surveys, or participatory design with developers, testers, and managers, could reveal insights about cognitive load, developer trust, integration challenges, and feedback loops. These insights are vital for shaping RL techniques that are not only performant but also accessible, interpretable, and aligned with existing tooling ecosystems (e.g., IDEs, CI/CD pipelines, testing frameworks).

\textcolor{teal}{\faSmileO}~\textbf{Opportunity 2 - Exploring RL-for-SE Coverage Throughout the SE Lifecycle.}
The observed gap suggests a great opportunity for future RL-for-SE research lies in broadening its scope beyond the current emphasis on test generation within \textbf{software quality assurance} (SQA).
While test generation naturally aligns with RL's exploration capabilities, SQA encompasses many other subtasks where RL can provide value. These include fault localization (e.g., identifying the root cause of test failures based on runtime signals), mutation testing (e.g., deciding which code mutations to evaluate), regression testing optimization, and test oracle generation. 
Beyond SQA, RL offers underexplored potential in \textbf{software maintenance}, which includes tasks such as automated code review and program repair. 
For instance, RL agents can learn to suggest context-aware code improvements during review by observing reviewer behavior, or to prioritize and apply security patches based on past vulnerability fix patterns and severity signals for security code review. 
In program repair, RL can explore and optimize over fix candidates, using reward signals based on compilation success, test suite performance, or human-in-the-loop feedback.
In \textbf{software development}, RL can guide code synthesis by incrementally constructing solutions while receiving feedback from compilation results or code quality metrics. It can also improve code completion and refactoring by learning from developer interaction histories and best practices. Moreover, RL-driven development assistants can adaptively recommend APIs or coding patterns tailored to project context and developer behavior.
While currently rare, \textbf{software design} presents another valuable application area. RL can assist in architectural decision-making by balancing trade-offs such as modularity, latency, scalability, and cost in evolving systems. 
For example, it could optimize microservice decompositions or dynamic component selection in reconfigurable systems. 
In \textbf{software requirements engineering}, RL could be used to refine and prioritize requirements dynamically based on stakeholder interactions, usage patterns, or evolving constraints, especially when combined with LLMs to interpret natural language inputs.
Finally, \textbf{software management} also remains vastly unexplored.
RL can be applied to backlog prioritization, sprint planning, team assignment, or resource scheduling, where agents learn from historical data to improve project outcomes over time, showing potential for intelligent DevOps.

\textcolor{teal}{\faSmileO}~\textbf{Opportunity 3 - Organic Combination of LLMs and RL.} A promising opportunity for advancing RL-for-SE research lies in the organic combination of RL and LLMs, a paradigm that unifies the strengths of both learning systems to address complex SE tasks more effectively. 
While RL excels at sequential decision-making and optimization under uncertainty, LLMs offer powerful language understanding, generation, and knowledge integration capabilities. 
Their synergy can enable adaptive, context-aware agents capable of performing sophisticated SE tasks such as automated code repair, test case generation, and commit message synthesis with minimal human intervention. 

Among the 115 surveyed papers, we observed that since 2024, a growing subset (10 studies) has begun to explore the integration of LLMs and RL to address SE challenges. 
Our manual analysis reveals that this hybrid approach is currently applied to a limited range of SE tasks, including code generation (3 papers), test generation (3 papers), bug localization (1), vulnerability detection (1), fuzzing (1), and code review (1).
This trend signals a promising and timely opportunity for the SE community to further investigate how LLM-guided RL or RL-enhanced LLMs can enable more intelligent, interpretable, and robust automation across diverse phases of the software development lifecycle. 
In particular, Reinforcement Learning from Human Feedback (RLHF)~\cite{christiano2017deep} holds unique potential for SE due to its ability to align model behaviors with nuanced human preference, an essential requirement in many SE tasks where correctness is necessary but not sufficient, such as comment generation, code review, recommendation tasks which involve trade-offs in readability, maintainability, and performance.
Moreover, RLHF can mitigate risks in safety-critical SE domains, such as vulnerability detection or program repair, by incorporating human feedback loops that penalize unsafe or non-robust behaviors during training. 

\textcolor{teal}{\faSmileO}~\textbf{Opportunity 4 - Toward Unified Open-Source Data and Evaluation Benchmarks.} 
The current landscape of RL-for-SE research reveals a pressing need for the construction of standardized open-source data and evaluation benchmarks. 
To bridge this gap, the SE community stands to benefit from benchmark initiatives that provide both diverse datasets and comprehensive evaluation protocols. These benchmarks should not only include high-quality, task-specific environments suitable for RL (e.g., test generation, code repair, vulnerability mitigation) but also reflect realistic constraints such as reward sparsity, partial observability, and delayed feedback. 
Furthermore, we advocate for greater methodological transparency and accessibility: researchers should clearly document the specific RL algorithms employed and release comprehensive replication packages, including datasets, training scripts, and configuration files. Such efforts are essential to foster cumulative progress and accelerate innovation in RL-for-SE.
Simultaneously, establishing unified evaluation frameworks for certain SE tasks that balance effectiveness and efficiency metrics will improve reproducibility and enable fair comparisons across techniques, and foster real-world adoption.

\textcolor{teal}{\faSmileO}~\textbf{Opportunity 5 - From Custom Hacks to Core Innovations: Unlocking RL's Potential in SE.}
While existing RL-for-SE research has made significant progress in addressing task-specific challenges, there is a noticeable opportunity to shift focus toward RL architecture design rather than solely optimizing task-specific applications. 
The majority of current studies emphasize adapting RL to practical SE tasks through custom optimizations, such as dataset construction or heuristic-based reward engineering, with limited attention to advancing the core RL architectures. 
For example, only a small percentage of studies explore novel state representations, action space designs, or reward modeling that can generalize across diverse SE tasks. 
By prioritizing innovations in RL conceptual design (e.g., modular RL architectures, transferable state-action abstractions, or scalable reward mechanisms), future research can enable broader applicability and improve the adaptability of RL techniques beyond narrowly defined problems. 
Such advancements would not only enhance the theoretical foundations of RL-for-SE but also bridge the gap between academic research and real-world industrial applications.

\textcolor{teal}{\faSmileO}~\textbf{Opportunity 6 - Development and Industrial Evaluation of RL-Based Software Engineering Tools.}
While the majority of RL-for-SE research focuses on conceptual frameworks or algorithmic innovations, very few studies have materialized as deployable tools, and even fewer have undergone systematic evaluation in industrial environments. 
This presents a significant opportunity to transition from academic prototypes to practical, tool-supported solutions that can be embedded into real-world software engineering workflows.
Moreover, collaboration with industry partners would provide access to realistic data, complex software artifacts, and feedback from practitioners, enabling iterative refinement of the RL models. 
This real-world grounding could help address issues such as reward misalignment, data drift, or inference latency that are rarely captured in lab-scale evaluations.
Ultimately, this direction would significantly elevate the impact and credibility of RL-for-SE by demonstrating its effectiveness in practice, uncovering hidden challenges, and driving future research to solve them. 
It also aligns with the increasing demand from industry for intelligent, adaptive, and explainable automation in software development processes.

\begin{tcolorbox}[colback=gray!5,colframe=awesome,title= RQ5 Summary]
\textcolor{BurntOrange}{\faStar}~Based on our literature review, we identify a set of key challenges and outline several actionable opportunities to advance RL-for-SE research. 
These insights span across research scope expansion, data availability and diversity, methodological innovations, and evaluation practices.
\end{tcolorbox}
\section{Threats to Validity}
\label{sec:threats}
We now discuss the potential threats to validity of this survey, focusing on external, internal, construct, and conclusion validity concerns.

\textbf{External Validity.} 
The generalizability of our findings may be limited by the scope of the selected venues and time frame. 
While our review focused on 22 premier software engineering conferences and journals, it is possible that relevant studies published in other venues or emerging workshops were excluded. 
However, we believe that the selected venues represent the core of high-quality research in the field, and the insights drawn from these works are sufficiently representative of the current state of RL applications in software engineering.
Additionally, our dataset is limited to papers published from 2015 onward, which may overlook foundational or exploratory work prior to the deep RL era.

\textbf{Internal Validity.} 
One threat to internal validity lies in the selection and screening process of the surveyed papers. 
Despite our systematic methodology, there is a risk of inadvertently omitting relevant studies due to limitations in search terms, database coverage, or manual filtering. 
To mitigate this threat, we carefully designed our inclusion and exclusion criteria and conducted a manual screening of the retrieved papers by two authors of this study. Additionally, to enhance the completeness of our survey, we applied a snowballing process to identify and include relevant studies that may not have been captured in the initial search.

\textbf{Construct Validity.} A key threat to construct validity is the definition and manual categorization of various research aspects, such as software engineering topics, tasks, and RL algorithms.
The boundaries between some categories (e.g., development vs. maintenance, or Q-learning vs. policy-gradient methods) may be ambiguous, leading to potential misclassification. 
To mitigate this, we referred to established taxonomies where available and cross-validated labels among multiple reviewers to reduce subjective bias.
For classification dimensions lacking an established taxonomy, we adopted a systematic open-coding process, including card sorting. 
This process involved collaborative labeling through a round-table discussion between two authors to ensure consistency and reliability in the categorization.

\textbf{Conclusion Validity.} The interpretation of trends and insights is another potential threat. 
Some observed patterns (e.g., growth in RL-for-SE studies or popularity of certain algorithms) may be influenced by external factors such as broader ML/AI trends, rather than inherent advantages of the approaches themselves. 
We have taken care to present quantitative findings objectively and supported our claims with multiple data points.

\section{Conclusion}
This work performed an SLR on 115 RL-for-SE papers published in 22 premier software engineering conferences and journals since the introduction of deep reinforcement learning in 2015. 
To comprehensively investigate the application of RL techniques in software engineering, we formulated five research questions that investigate various aspects. 
Our trend analysis in RQ1 reveals a clear upward trajectory in the number of RL-for-SE papers published annually, particularly since 2022. Additionally, the majority of the surveyed studies focus on proposing novel techniques or methodologies.
The SE research topic analysis in RQ2 indicates that RL-for-SE research is heavily concentrated in Software Quality Assurance, followed by Software Development and Software Maintenance.
The findings from RQ3 suggest that existing studies show a strong preference for model-free, value-based methods. 
At the same time, an emerging trend toward the adoption of more advanced RL strategies, such as actor-critic methods, has also been observed.
RQ4 reveals strengths and limitations in current research practices, including data usage, model design and optimization techniques, evaluation metrics, and replicability.
Last, we outlined a set of current challenges and potential opportunities to better guide future research on applying RL in SE.

\section*{Acknowledgment}
This work was supported by the National Key Research and Development Program of China (Grant No. 2024YFB4506300), the National Natural Science Foundation of China
(Grant Nos. 62322208, 62232001, 62472310).

\section*{Declarations}
\subsection*{Funding and/or Conflicts of interests/Competing interests}
All authors certify that they have no affiliations with or involvement in any organization or entity with any financial interest or non-financial interest in the subject matter or materials discussed in this manuscript.

\bibliographystyle{ACM-Reference-Format}
\bibliography{main}










\end{document}